\documentclass[sigconf]{acmart}

\copyrightyear{2025}
\acmYear{2025}
\setcopyright{acmlicensed}
\acmConference[C\&C '25]{Creativity and Cognition}{June 23--25, 2025}{Virtual, United Kingdom}
\acmBooktitle{Creativity and Cognition (C\&C '25), June 23--25, 2025, Virtual, United Kingdom}
\acmDOI{10.1145/3698061.3726915}
\acmISBN{979-8-4007-1289-0/2025/06}

\newcommand{\killpunct}[1]{}

\begin{document}

\title{Fuzzy Linkography: Automatic Graphical Summarization of Creative Activity Traces}

\author{Amy Smith}
\affiliation{%
  \institution{Queen Mary University of London}
  \city{London}
  \country{United Kingdom}}
\email{amy.smith@qmul.ac.uk}

\author{Barrett R. Anderson}
\affiliation{%
  \institution{Independent Researcher}
  \city{San Mateo}
  \state{California}
  \country{USA}}
\email{barrettrees@gmail.com}

\author{Jasmine Tan Otto}
\affiliation{%
  \institution{Independent Researcher}
  \city{San Mateo}
  \state{California}
  \country{USA}}
\email{ottojasmine@gmail.com}

\author{Isaac Karth}
\affiliation{%
  \institution{Independent Researcher}
  \city{Berkeley}
  \state{California}
  \country{USA}}
\email{isaac@isaackarth.com}

\author{Yuqian Sun}
\affiliation{%
  \institution{Midjourney}
  \city{London}
  \country{United Kingdom}}
\email{ysun@midjourney.com}

\author{John Joon Young Chung}
\affiliation{%
  \institution{Midjourney}
  \city{San Francisco}
  \state{California}
  \country{USA}}
\email{jchung@midjourney.com}

\author{Melissa Roemmele}
\affiliation{%
  \institution{Midjourney}
  \city{San Francisco}
  \state{California}
  \country{USA}}
\email{mroemmele@midjourney.com}

\author{Max Kreminski}
\affiliation{%
  \institution{Midjourney}
  \city{San Francisco}
  \state{California}
  \country{USA}}
\email{mkreminski@midjourney.com}
\orcid{0009-0002-6268-4033}

\renewcommand{\shortauthors}{A. Smith, B.R. Anderson, J.T. Otto, I. Karth, Y. Sun, J.J.Y. Chung, M. Roemmele, and M. Kreminski}

\begin{abstract}
Linkography---the analysis of links between the \emph{design moves} that make up an episode of creative ideation or design---can be used for both visual and quantitative assessment of creative activity traces. Traditional linkography, however, is time-consuming, requiring a human coder to manually annotate both the design moves within an episode and the connections between them. As a result, linkography has not yet been much applied at scale. To address this limitation, we introduce \emph{fuzzy linkography}: a means of automatically constructing a linkograph from a sequence of recorded design moves via a ``fuzzy'' computational model of semantic similarity, enabling wider deployment and new applications of linkographic techniques. We apply fuzzy linkography to three markedly different kinds of creative activity traces (text-to-image prompting journeys, LLM-supported ideation sessions, and researcher publication histories) and discuss our findings, as well as strengths, limitations, and potential future applications of our approach.
\end{abstract}


\begin{teaserfigure}
\centering
  \includegraphics[width=\textwidth]{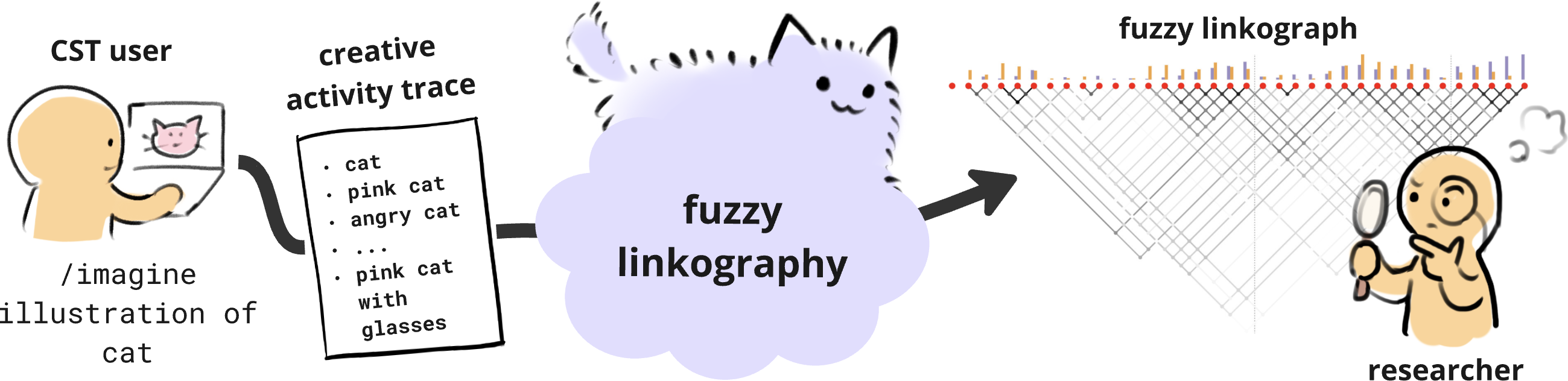}
  \caption{Fuzzy linkography allows for the rapid translation of user activity logs from digital creativity support tools (and other traces of creative activity) into rough graphical summaries, suitable for visual and quantitative inspection by researchers. 
  }
  \Description{A transcript of a design or ideation session passes into the process, and a fuzzy linkograph emerges from the other side. A curious researcher holds up a magnifying glass to inspect the patterns in the links between the design moves making up the backbone of the linkograph.}
  \label{fig:teaser}
\end{teaserfigure}

\maketitle

\section{Introduction}
Digital creativity support tools (CSTs)~\cite{ShneidermanCSTs,CSTReviewChung,CSTReviewFrich}---software systems intended to support users' creative activity---have become increasingly prevalent in the last few decades, both in professional and casual contexts~\cite{CasualCreators}. Recent years have seen especially strong growth in the adoption of AI-based CSTs---in creative domains as wide-ranging as writing~\cite{DSIIWA}, visual art~\cite{ThePromptArtists,IsWritingPromptsMakingArt,AIReadymades}, game design~\cite{FriendCollaborator,Germinate}, fashion design~\cite{FashionQ}, worldbuilding~\cite{Patchview}, and music~\cite{CoCoCo}.

Widespread use of digital CSTs permits the automatic capture of traces of human creative activity at a previously unprecedented scale, which could be of assistance both in developing theories of creativity and in evaluating CSTs. Evaluating CSTs is difficult~\cite{EvaluatingCSTs} and has been known to be difficult since the first days of CST research~\cite{EvaluatingCSTsWorkshopReport}; many evaluation approaches focus on the assessment of CST users' subjective experiences of the creative process~\cite{CSI,MICSI}, while others focus more on the products of CST use~\cite{ExpressiveCommunication,HomogenizationEffects}. However, recording of user interaction traces also permits the evaluation of \mbox{(co-)creative} \emph{interaction dynamics}, including via the visualization and analysis of user trajectories through design space~\cite{DrawingApprenticeInteractionDynamics,DesignStyleClustering,ERaCA,Drawcto}. These evaluations are process-focused and attempt to characterize how creative activity evolves over time, allowing them to directly inform process models of creativity~\cite{ModelsCreativeProcess,ProcessDefinition}.

A similarly visual, temporal, and process-focused method of assessing creative activity---\emph{linkography}---has been employed since the early 1990s~\cite{Linkography,LinkographyOriginal} in domains ranging from architecture~\cite{ArchitectureLinkography} and product design~\cite{ProductDesignLinkography} to animation~\cite{EntropyOfLinkography} and sound design~\cite{SoundDesignLinkography}. Linkography hinges on annotation and visualization of the \emph{design moves} that make up an episode of creative activity and the \emph{links} between related moves; various statistics can also be computed on the resulting graphs~\cite{Linkography,LinkographyGero}. Although linkography is sometimes employed to investigate user interactions with CSTs~\cite{CADLinkography}, including AI-based CSTs~\cite{PromptLinkography}, it remains a ``costly research method, both in terms of time and resources''~\cite{LINKOgrapher}; as a result, linkography is still rarely used to analyze large numbers of creative activity traces captured by digital CSTs, and some researchers have even declined to employ linkography at smaller scales due to its ``logistical and labour overheads''~\cite{DecliningLinkography}.

To address this limitation, we introduce \textbf{fuzzy linkography}: a technique for automatically producing (imperfect) linkographs from sequences of recorded design moves, using a computational model of semantic similarity as a stand-in for the human annotator of links between design moves. We apply our approach to three markedly different kinds of creative activity traces: text-to-image prompting journeys, LLM-supported ideation sessions, and researcher publication histories. Across these domains, we identify familiar patterns from manually constructed linkographs in fuzzy linkographs; highlight recurring motifs in image prompting journeys that fuzzy linkographs make visible; examine how users integrate AI-generated material during co-creative ideation; and probe the use of fuzzy linkography to enable low-cost experimentation with linkographic analysis in new research contexts. Finally, we discuss strengths and limitations of our approach, as well as new applications of linkography that may be unlocked by rapid and automatic construction of linkographs at scale.

\section{Linkography: A Brief Primer}
Traditional linkography has been explicated in a number of prior publications (e.g., \cite{Linkography,LinkographyGero,LINKOgrapher,UsingLinkography,PromptLinkography}), so we do not attempt to describe it fully here. Instead, we first present a brief overview of key terms and concepts from the linkography literature. Then we discuss (in Section \ref{sec:FuzzyLinkography}) how we update these concepts for application to automatically constructed linkographs that have continuously weighted rather than binary links.

A linkograph represents a single \emph{design episode} and consists of two primary components: a sequence of \emph{design moves} and a set of pairwise \emph{links} between these design moves. \textbf{Design moves} represent concrete changes made to the \emph{design situation}, and are conventionally plotted left to right, with each move represented by a single dot and uniform spacing between moves (regardless of how much time elapsed between each pair of moves). \textbf{Links} represent human-annotated connections between related design moves: if a later design move can be said to \emph{build on} an earlier design move, a link is drawn connecting these two moves. Links are conventionally drawn below the move sequence.

Several statistics can be computed on linkographs. A move's \textbf{forelink count} is defined as the number of links between this move and any later moves to which it is connected; this is conventionally interpreted as indicating the extent to which a move \emph{influenced} later moves. A move's \textbf{backlink count} is defined similarly, as the number of links between this move and any \emph{previous} moves to which it is connected; this is seen as indicating the extent to which a move \emph{integrates} several prior moves. Moves with an especially high link count in either direction are known as \textbf{critical moves}; forelink critical moves are sometimes interpreted as moments of useful \emph{divergence} (i.e., divergence that is later built on repeatedly), while backlink critical moves are sometimes interpreted as moments of useful \emph{convergence} (i.e., bringing together several ideas from earlier in the design episode). Finally, a \textbf{link density index} (LDI) for the linkograph as a whole can be calculated by dividing the total number of links by the total number of moves; this value indicates the overall interconnectedness of the moves in the graph, with a higher LDI suggesting that moves are more strongly related to one another overall.

To quantify the unpredictability of links in a graph (as a rough proxy for the dynamism of the whole design episode), several measures of \textbf{link entropy} are also sometimes employed. Forelink and backlink entropy are first computed on a move-by-move basis, by determining the probability that a link does or does not exist between a given move $M$ and any following (forelinkable) or preceding (backlinkable) move; these values are then summed across all moves to give a total forelink and backlink entropy for the whole graph. \textbf{Horizonlink entropy} is a similar measure, but quantifies the unpredictability of links at each possible \emph{horizon level}, i.e., each possible distance between pairs of moves; horizonlink entropy is first calculated for the set of all move pairs that are exactly \emph{one} move apart, then for the set of all move pairs that are \emph{two} moves apart, and so on, and these values are again summed together to give a total horizonlink entropy for the whole graph. Finally, the three different graph-level entropy values (forelink, backlink, and horizonlink) can themselves be summed to give an \textbf{overall link entropy} for the graph as a whole.

\begin{figure*}
    \centering
    \includegraphics[width=\linewidth]{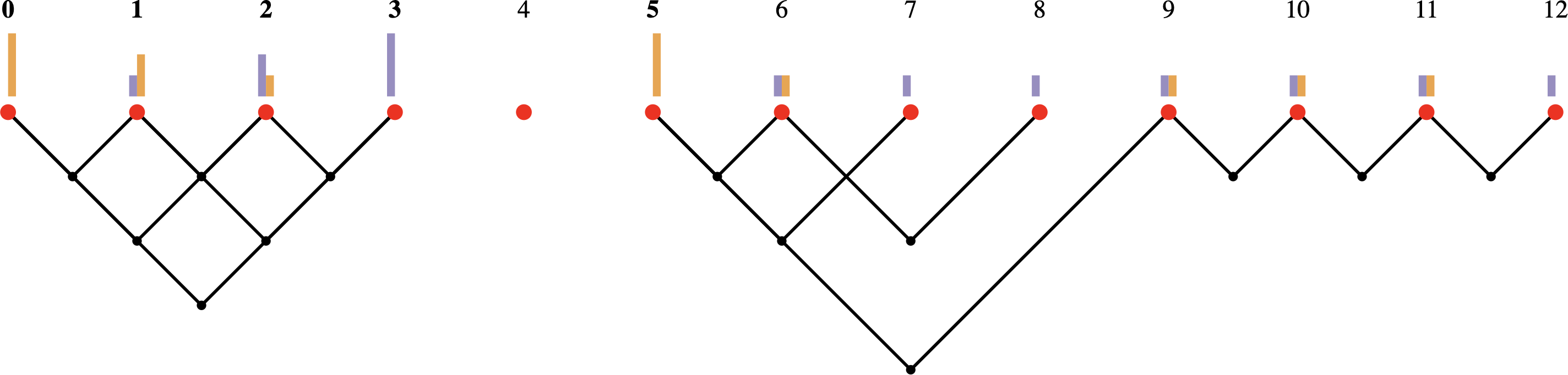}
    \caption{Standard linkographic patterns. Moves 0--3 form a \emph{web}; move 4 is an \emph{orphan}; moves 5--9 form a looser \emph{chunk}; and moves 9--12 form a \emph{sawtooth}. Above each move, purple and orange bars indicate the move's \emph{backlink} and \emph{forelink count} respectively; moves with an especially high link count in either direction may be deemed \emph{critical moves}, e.g., the \emph{forelink critical move} 5.}
    \Description{A sequence of thirteen design moves is drawn from left to right as red dots, with each one labeled by a number (0--12) and related moves connected by lines drawn below the move sequence. Moves 0--3 are all connected to one another; move 4 isn't connected to anything; move 5 is connected to moves 6, 7, and 9; move 8 is connected to move 6; and moves 9--12 are only connected to one another, except for move 9's backlink to the cluster starting with move 5. A tall orange bar over moves 0 and 5 indicates that both of these moves are heavily forelinked, and a tall purple bar over move 3 indicates that this move is heavily backlinked; other moves also have shorter orange and purple bars above them, indicating their less extreme degrees of backward and forward connectivity.}
    \label{fig:PatternExamples}
\end{figure*}

Visually, linkographs are often analyzed in terms of the structures that appear within them. \citeauthor{Linkography} \cite{Linkography} describes three key pattern types that may be seen in linkographs (Figure \ref{fig:PatternExamples}): \textbf{chunks}, or sequences of interrelated moves that begin with a clear ``inciting'' move and are mostly linked back to that incident; \textbf{webs}, or more tightly interrelated clusters of moves in which each move is linked to almost every other; and \textbf{sawtooths}, or sequences of moves in which each move is related only to its immediate predecessor and successor (potentially suggesting the development of a single idea that is largely not tied into the rest of the design situation). Linkless moves are called \textbf{orphans} and indicate ignored digressions. \citeauthor{SaturatedForelinks} \cite{SaturatedForelinks} also suggest that a move with fully saturated forelinks (i.e., forelinks to every move that follows it) may be indicative of \emph{fixation} on a single idea~\cite{DesignFixation,FixationReviewGero,FixationReview2018,FixationReview2019} and thus a lack of forward progress in the episode past the forelink-saturated move.

\section{Fuzzy Linkography}
\label{sec:FuzzyLinkography}

\begin{figure*}
    \centering
    \includegraphics[width=0.8\linewidth]{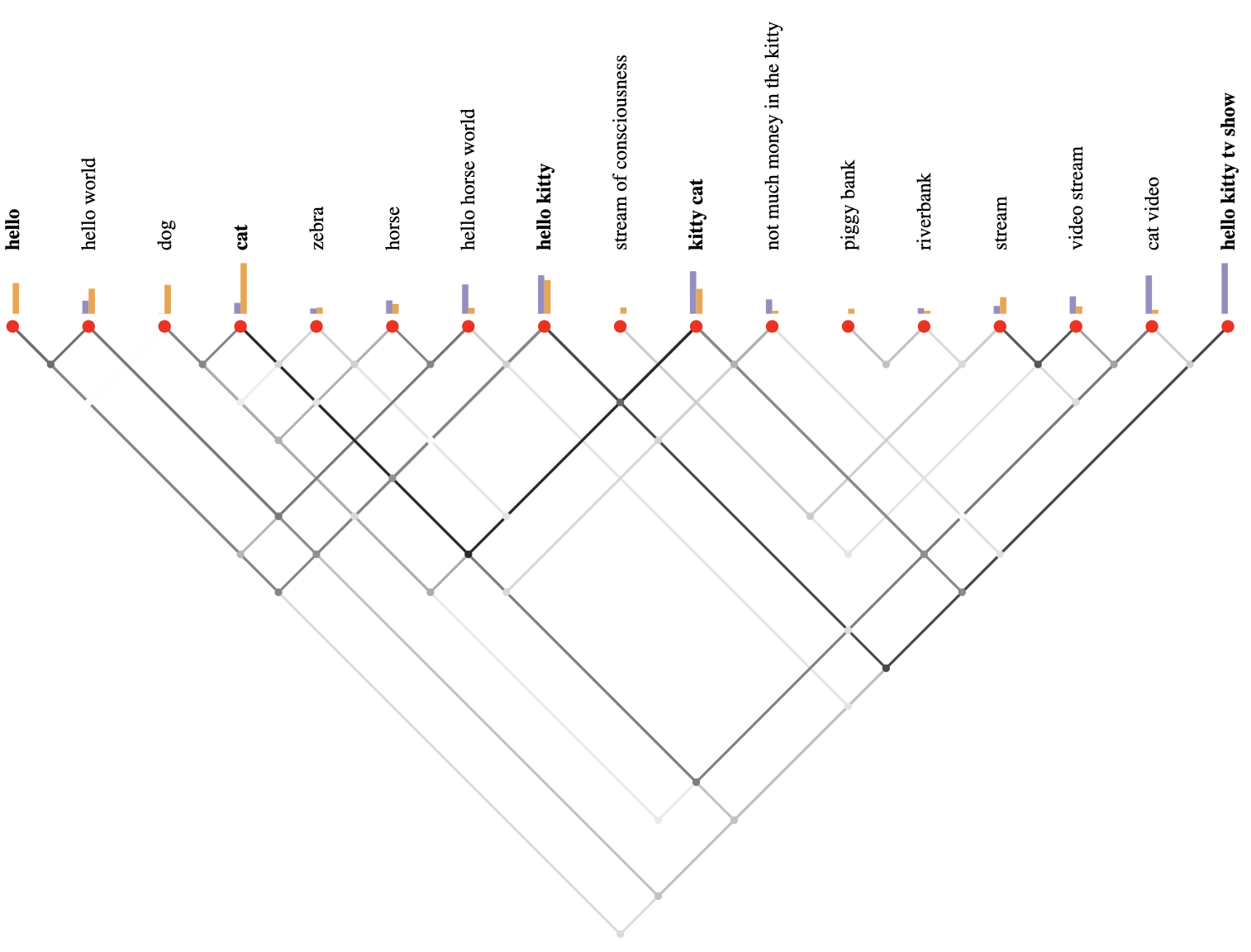}
    \caption{An example fuzzy linkograph of an open-ended stream-of-consciousness ideation activity. Links between moves are colored according to their strength (with darker lines indicating stronger links). Move text is displayed directly above the move markers. This graph features several critical moves: ``hello'' and ``cat'' are forelink critical moves that introduce the themes of greeting and cats respectively; ``hello kitty'' initially integrates these two themes; and ``hello kitty tv show'' is a backlink critical move, fully integrating these themes from the session's first half with the ``video'' theme from its second.}
    \Description{Another linkograph like the previous, but this one features links of different line weights, ranging from a very faint color for weak links all the way to a very high-contrast color for strong links. The graph consists of seventeen moves, each of which is labeled with a short text string; in order the move labels are "hello", "hello world", "dog", "cat", "zebra", "horse", "hello horse world", "hello kitty", "stream of consciousness", "kitty cat", "not much money in the kitty", "piggy bank", "riverbank", "stream", "video stream", "cat video", and "hello kitty tv show". Strong links are present between those moves that mention cats or kitties, with the exception of "not much money in the kitty", which features several weaker links to these cat-related moves instead. Greeting-related moves are linked to one another by medium-strength or strong links, and animal-related moves are linked to one another at different levels of strength. The development of the "stream" theme (and its gradual transition into the "video" theme) in the latter half of the graph is made visible by weak to medium-strength links between several relevant moves. The moves "hello kitty" and "hello kitty tv show" both seem especially strongly backlinked to a number of previous moves; these are important moments of convergence within the graph as a whole.}
    \label{fig:ExampleFuzzyLinkograph}
\end{figure*}

Fuzzy linkography (Figure~\ref{fig:ExampleFuzzyLinkograph}) is much like traditional linkography, with two key differences. First, links between moves are \emph{automatically inferred} by a computational process rather than manually annotated by a human coder. Second, links are represented as numbers ranging from zero to one (indicating the \textbf{strength} of semantic association between a pair of moves) rather than binary on/off values. Continuous link strength values, and their imperfect correlations with human assessments of move relatedness, are what give fuzzy linkographs their fuzziness; rather than forcing the machine annotator to make an authoritative-seeming binary choice about whether each pair of moves is or is not related, we prefer to pass information about move associations that the machine finds ambiguous or uncertain along to the human user of linkography.

In our implementation of fuzzy linkography, links between moves are established via an \emph{embedding model}~\cite{SentenceTransformers} that translates textual descriptions of design moves into vectors. For each pair of moves, we use cosine similarity between the embedding vectors representing each move to determine the strength of the link between these moves. A numeric similarity threshold $t$ is used to discard weak associations between moves, so that not all moves are judged as being linked to some extent; raw cosine similarity values exceeding this threshold are then linearly rescaled from the range $[t, 1]$ to the range $[0, 1]$ to establish link strengths.

When rendering fuzzy linkographs, we use link color---ranging from white, for very weak links, to black, for very strong links---to indicate the strength of each link. 
``Fuzzy'' lighter-colored links thereby serve to communicate model uncertainty about move association; visual ``fuzziness'' has been found to be an intuitive way to convey uncertainty~\cite{UncertaintyVisualization}, which can help to improve the transparency of analyses that rely on machine learning models~\cite{UncertaintyTransparency,UncertaintyDesignMaterial}.

Because fuzzy links are represented as numeric strength values rather than binary on/off states, the quantitative measures calculated in traditional linkography must be updated to work with fuzzy linkographs. Forelink and backlink weight values on a fuzzy linkograph can be calculated simply by summing the strengths of a move's forelinks and backlinks respectively. Link density index values can similarly be computed by summing the strengths of all links in the graph and dividing this by the total number of moves.

Entropy values can be calculated for fuzzy linkographs by treating the strength of a link between two moves (which already ranges from 0 to 1) as the \emph{probability} of a binary link existing between these same moves. In \citeauthor{LinkographyGero}'s original formulation of link entropy, $p(ON)$---the probability of a possible link existing---is calculated for each row of forelinks, backlinks, and horizonlinks (i.e., each ``state'' $s$) as the actual number of links that are present in this state divided by the maximum number of links $n_s$ that are possible in this state. We thus calculate $p(ON)_s$ for the same states as:

\begin{equation}
   p(ON)_s = \frac{\sum_{v \in L_s} v}{n_s}
\end{equation}

...where $L_s$ is the set of strength values $v$ of all links that are present in state $s$. We then follow \citeauthor{LinkographyGero}'s derivation~\cite{LinkographyGero} of $p(OFF)_s$ and all downstream entropy values.

One consequence of this interpretation is that a fuzzy linkograph in which every move is connected to every other by ambiguous links may be judged as having greater entropy than a linkograph with the same LDI, but an equal number of very strong and completely absent links. This is because the graph saturated with ambiguous links can be viewed as distributing uncertainty more evenly across the whole graph. Though we have not seen this cause any practical problems so far, it strikes us as somewhat counterintuitive. At any rate, we do not make evaluative use of entropy in the analyses we present here, because our goal is not to comparatively evaluate the success of different creative practices or tools but to broadly characterize creative behaviors observed in different situations; further refinement of entropy measures on fuzzy linkographs may be a good topic for future work.

To support widespread application of fuzzy linkography, we open-source the code we use to construct, visualize, and quantitatively analyze linkographs.\footnote{\url{https://github.com/mkremins/fuzzy-linkography} 
} All analyses that we report on in this paper assess design move similarity via the \texttt{all-MiniLM-L6-v2} sentence embedding model~\cite{SentenceTransformers}---an open-source, open-data and open-weights model that has previously been validated against a human baseline for assessment of semantic similarity in the context of open-ended ideation~\cite{HomogenizationEffects}. For all analyses reported in this paper, we use a fixed similarity threshold $t = 0.35$ as the minimum similarity score for which we infer a link; this value seems to work well with our chosen embedding model, although in the future, it may be worth investigating whether the value of $t$ can somehow be chosen in a more principled way.

\section{Analyzing Image Prompting Journeys}
As a first case study, we apply fuzzy linkography to the analysis of user image prompting journeys~\cite{AIOrMe} in a graphical creativity support tool built around a popular commercial text-to-image diffusion model. Image prompting practices have previously been analyzed computationally via topic modeling~\cite{T2IPromptAnalysis,T2IWhyAndHow,T2IExploratory}, and the dynamics of narrowly goal-directed prompting have been investigated at a large scale through natural language processing techniques~\cite{T2IThreads}, but we are unaware of any prior large-scale examination of user-level idea development in text-to-image prompting. Our dataset consists of 6,424 interaction traces documenting every image prompt submitted by every user of the tool during the period from Dec 11--31, 2024; each trace consists of the complete sequence of prompts submitted by a single user during this period. To exclude data from users who did not make much use of the tool, we filtered these interaction traces to 1,879 ``substantial'' traces that contain at least seven image prompts each. The longest trace in our filtered dataset is 536 moves long, and the filtered traces overall have a median length of 14.

We consider each interaction trace as a separate design episode and construct a fuzzy linkograph of the episode, treating individual textual prompts submitted by the user as design moves. Our (unoptimized) Python implementation of link inference, running in Python 3.9.6 on an Apple MacBook Air (M2, 2022), takes 752.944 seconds to compute link strengths for the filtered set of traces---roughly 0.4 seconds per trace. We then compute several linkographic statistics on each trace, including link density index; move-level forelink and backlink weights (which can be used to help identify critical moves); graph-level forelink, backlink, and horizonlink entropy values; and an overall link entropy value summing up the other entropies. 

\subsection{Recurring Linkographic Motifs}
By visually inspecting the linkographs of each trace, noting recurring patterns, and investigating the specific prompts that are involved in these patterns, we can begin to build a taxonomy of structural \emph{motifs} that frequently appear in image prompting traces.

\subsubsection{Refinement Webs}
\begin{figure}
    \centering
    \includegraphics[width=\linewidth]{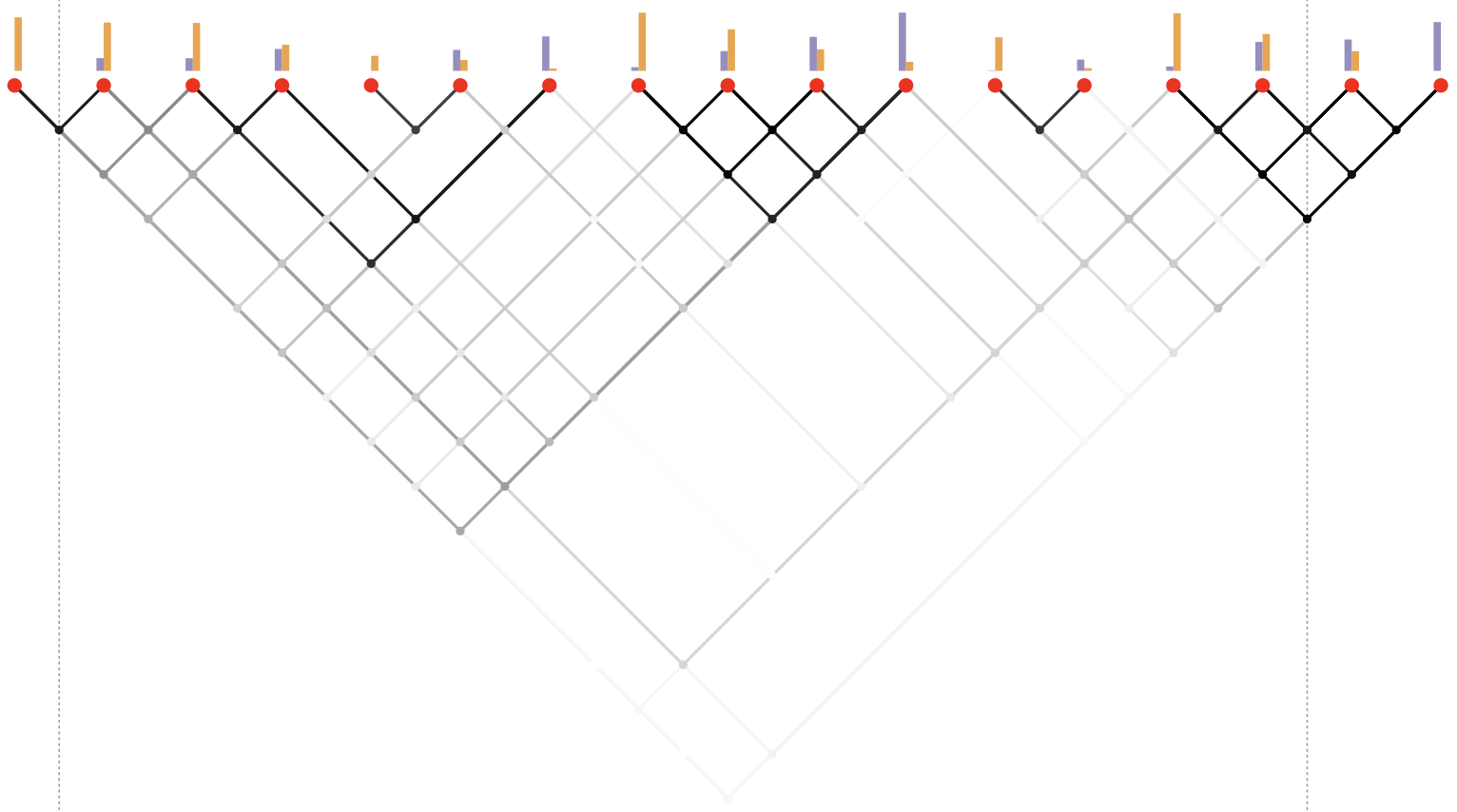}
    \caption{This image prompting linkograph contains three refinement webs: one at the beginning, one in the middle, and one at the end. The first web consists of mixed medium-strength and high-strength links, indicating a greater degree of prompt variation at an earlier level of conceptualization; the other two webs mostly consist of very small permutations of a relatively fixed prompt.}
    \Description{Another linkograph, this time with clear black lines indicating strong links; gray lines indicating moderate links; and faint lines indicating weak links. Most of the moves are connected strongly to immediate neighbors up to 2 or 3 moves away in tight little web patterns, but no further; there's only a handful of moderate-strength mid-range links and one or two weak long-range links.}
    \label{fig:RefinementWebs}
\end{figure}

The most common motif in our image prompting linkographs is a ``web'' of tightly interconnected moves, signifying a moment of prompt \emph{refinement}: the user gradually testing smaller and smaller variations on a prompt with relatively fixed subject matter as they narrow in on the images they want. Links within this web can often be seen to get gradually stronger from the beginning to the end of the sequence, as the submitted prompts become steadily more similar to one another---eventually often culminating in several retries of the exact same prompt. In Figure \ref{fig:RefinementWebs}, for example, three refinement webs correspond to a user's attempts to visualize an alien civilization; a particular location on that civilization's home planet; and members of a specific faction within that civilization, respectively. These webs seem to correspond roughly to the prompting ``threads'' identified by \citet{T2IThreads}.

Interestingly, we found that large webs of tightly interconnected moves are fairly common in the text-to-image linkographs we examined---contrasting somewhat with an earlier finding that a prompting-based interface usually resulted in relatively small webs compared to a sketching-based interface~\cite{PromptLinkography}. We suspect this may occur because the CST we studied generates multiple different images in response to a single prompt submission, perhaps increasing the perceived value to users of retrying identical or near-identical prompts to stochastically explore a space of possible results.

\subsubsection{Curiosity Zigzags}
\begin{figure}
    \centering
    \includegraphics[width=\linewidth]{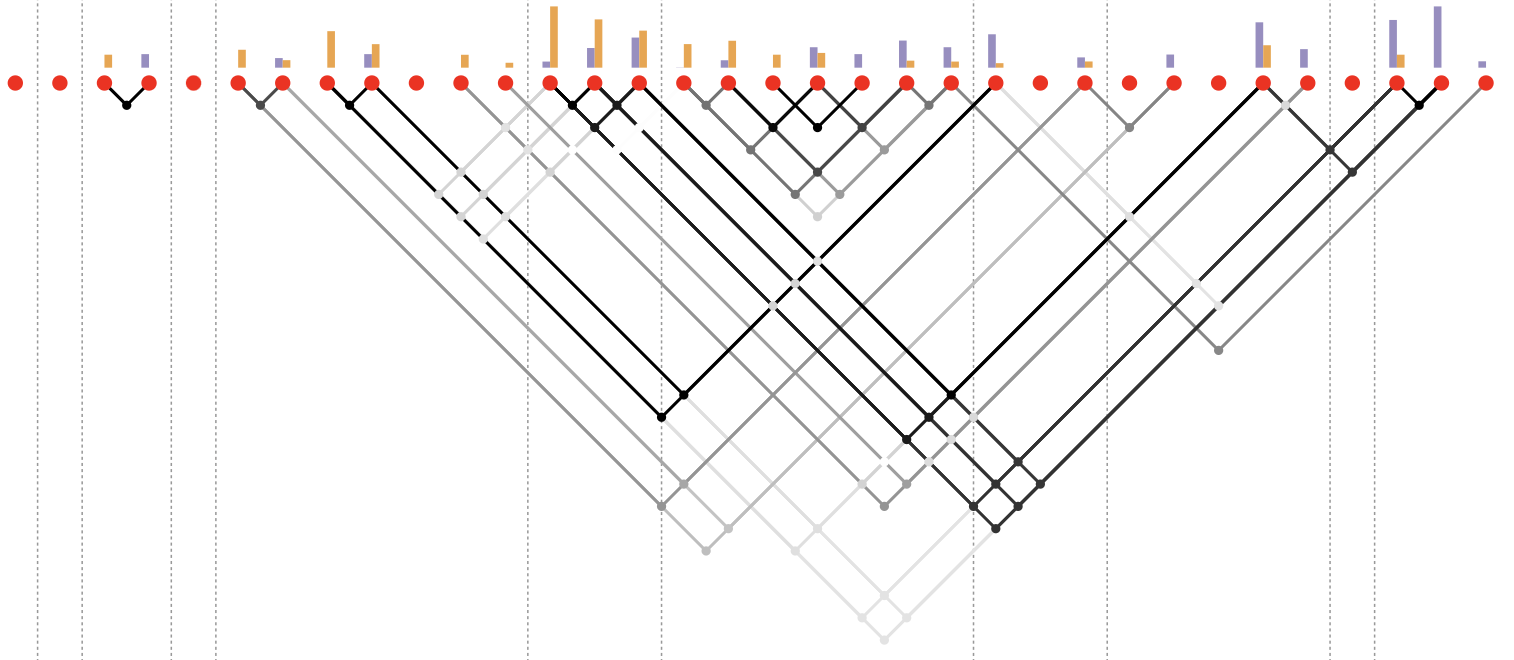}
    \caption{This image prompting linkograph demonstrates a curiosity zigzag, with relatively strong long-range links indicating returns to a central theme and disconnected chunks between these moves indicating periods of exploration.}
    \Description{Another linkograph, with "zigzag-shaped" strong links between some pairs of moves that are relatively far apart in time. Nested inside one of the zigzag shapes is a "chunk" of moves that are strongly linked to some of their closer neighbors, but this chunk is disconnected from the zigzag.}
    \label{fig:CuriosityZigzag}
\end{figure}

Another common motif in our image prompting linkographs is a large-scale ``zigzag'' structure temporally interspersed with smaller and largely unrelated chunks or webs, often corresponding to user alternation between a ``central theme'' that they keep returning to and a set of further-flung explorations of different ideas or themes. Some explorations seem to follow the permutation-based exploratory prompting dynamics observed by \citet{T2IExploratory}, while others exhibit greater variation within exploratory chunks. In Figure \ref{fig:CuriosityZigzag}, for instance, a user periodically returns to a single central subject---a cyclopean ``black spirit'' character---between largely unrelated and wide-ranging explorations of other themes.

This mirrors a pattern observed in an earlier visualization-based study of user trajectories through a casual CST's design space~\cite{ERaCA}, in which some users visibly alternated between outputs sampled from a relatively narrow ``home base'' of outputs and further-flung explorations that push the boundaries of the CST's expressive range~\cite{ExpressiveRange}. Relative lack of integration of further-flung explorations may indicate boundary-pushing, probing the CST's capabilities, or exploration for its own sake (perhaps indicative of the ``curious users of casual creators'' phenomenon identified by \citet{CuriousUsers}); in some cases we believe this pattern may also indicate the user's non-development of a coherent goal.

\subsubsection{Zigzags Toward Convergence}
\begin{figure*}
    \centering
    \includegraphics[width=0.8\linewidth]{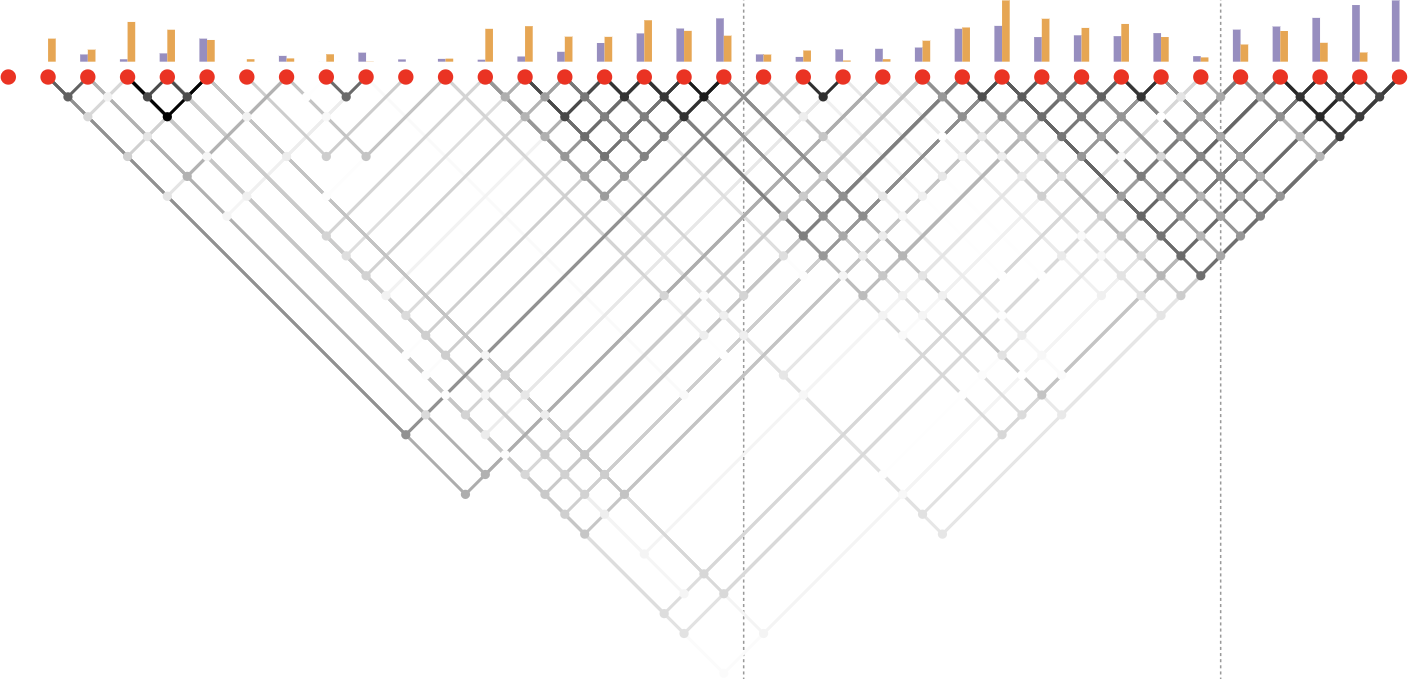}
    \caption{This image prompting linkograph's ``converging zigzag'' shape combines the \emph{refinement webs} motif with the \emph{curiosity zigzag} motif: alternation between a central theme of interest and more divergent explorations persists throughout the episode, but previously distinct themes are also brought together into a single image concept on two distinct occasions, each time resulting in a distinctive refinement web.}
    \Description{Another linkograph, containing two big and obvious "web" clusters of densely interconnected moves and a number of longer-range "zigzag" links of moderate strength. Unlike in the previous linkograph, the zigzag gradually becomes more entangled with the web structures over the course of the whole move sequence.}
    \label{fig:ConvergingZigzag}
\end{figure*}
One recurring motif combination, which we term the ``converging zigzag'', involves a curiosity zigzag with gradually growing integration between the ``main theme'' and divergent exploration chunks or webs over time (Figure \ref{fig:ConvergingZigzag}). This structure seems to be especially characterized by refinement webs of progressively strengthening links at the ``tail end'' of a move sequence, corresponding to integrations of previously unrelated themes.

Figure~\ref{fig:ConvergingZigzag} shows two such webs. The first web corresponds to a gradual drawing-together of two recurring \emph{subject} themes in this user's prompts (an insect and a ``cute old man'' character); the second corresponds to a drawing-together of this hybrid character subject with a visual \emph{style} that the user had previously explored in other, unrelated images.

Interestingly, the first refinement web occurs just before the user stops engaging with the CST for a while. The second refinement web also occurs near two temporal ``session breaks''; it crosses over the first break and terminates with the second, indicating that the user did not engage further with the CST following their completion of the second web. This seems to indicate that the user twice gradually formed a specific goal and progressed toward its realization, then stepped away from the CST once the goal was realized. More broadly, across our image prompting data, refinement webs toward the tail end of a user activity period generally seem to indicate the user forming and realizing a concrete goal, then stepping away with the ``final'' generated images.

\subsubsection{Variations in Temporal Structure}
\begin{figure*}
    \centering
    \begin{minipage}{0.48\textwidth}
        \centering
        \includegraphics[width=\linewidth]{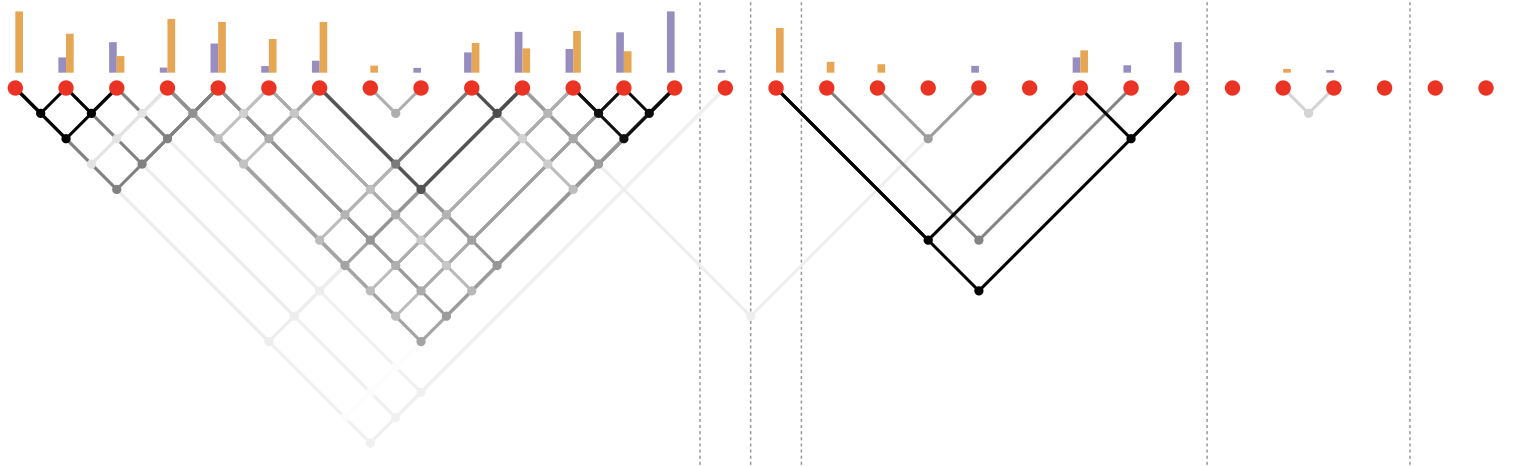}
        \caption{This image prompting linkograph shows clear semantic divisions between the user's focal themes during several significant temporal ``sessions'' of prompting. (Vertical dotted lines between moves indicate the user stepping away from prompting for at least 30 minutes, the same segmentation threshold applied by \citet{T2IExploratory}.) The first and longest session ends with a clear refinement web.}
        \Description{Another linkograph, with a big and obvious "chunk" of interrelated moves toward the start of the move sequence; a sparser "chunk" in the middle; and a tail of almost entirely unconnected moves at the end. The first two chunks are separated from one another by a pair of vertical dotted lines indicating significant temporal breaks in the move sequence; another such divider separates the second chunk from the tail, and the tail itself is split in two by another divider.}
        \label{fig:TemporalSessions}
    \end{minipage}
    \hfill
    \begin{minipage}{0.48\textwidth}
        \centering
        \includegraphics[width=\linewidth]{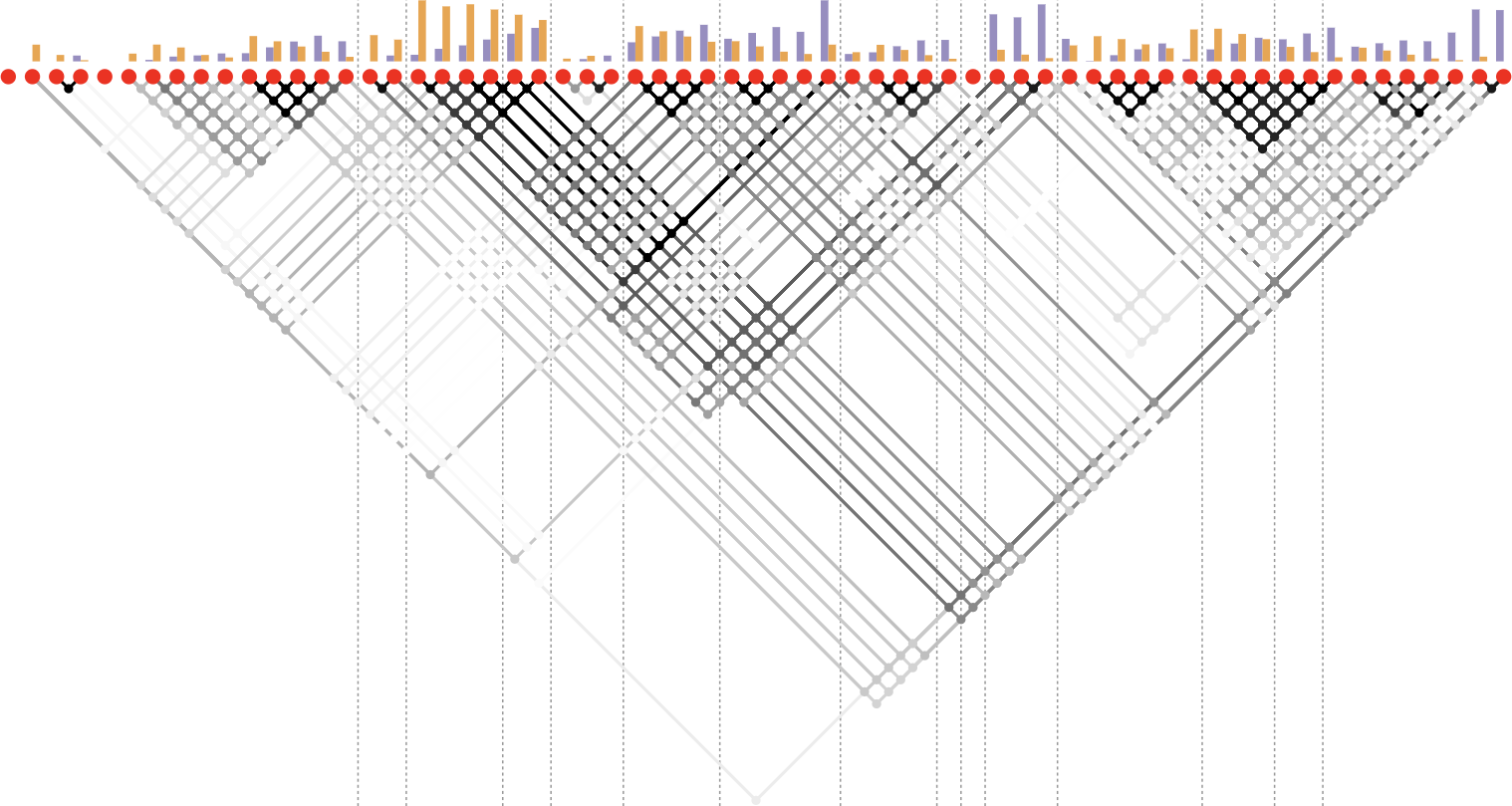}
        \caption{This image prompting linkograph, on the other hand, shows a user's focal themes persisting across multiple clear temporal divides in prompting activity. The user seems to be engaged in a long-running ``project''.}
        \Description{A linkograph of a long move sequence with many strong short-range and medium-range links, plus a good number of medium-strength long-range links. The overall impression is of a single giant chunk of interconnected moves. Many vertical dotted lines cut through the chunk, seemingly almost at random; they do not seem to break up the semantic connectivity of the whole move sequence.}
        \label{fig:NonTemporalProjects}
    \end{minipage}
\end{figure*}

We observed clear differences between users in terms of activity changes across breaks from prompting. The two most obvious temporal patterns we noticed in linkograph structure are temporally-divided \emph{sessions} (Figure \ref{fig:TemporalSessions}) and temporally non-divided \emph{projects} (Figure \ref{fig:NonTemporalProjects}). Sessions are chunks of related prompting activity that are cleanly bounded by the user taking a break from prompting on each side; projects are chunks of related prompting activity that span multiple such breaks.

We were initially surprised to see clear persistence of prompting subjects across temporal divides in significant numbers of traces, because in our anecdotal experience, the subjects of prompting are often drawn relatively directly from recent real-life influences. However, the simultaneous existence of users who frequently revisit subjects and users who do not seems to line up well with observations of diverse user goals and motivations in prior studies of text-to-image prompting~\cite{AIOrMe}. We also suspect that users of the CST we studied may be more frequently influenced by their own past prompts than users of other text-to-image tools, because the CST we studied presents a canvas-based interface in which past prompts and images may more readily persist as influences on a returning user. In other words, many text-to-image tools present users with a ``blank canvas'' (i.e., an empty text box) every time they begin a session of prompting---but a GUI-based tool that immediately presents users with outputs of their prior prompting as ``ingredients'', and allows direct remixing of these ingredients, may lead more users to undertake longer-term prompting projects.

\subsection{Trace Clustering}
\begin{figure*}
    \centering
    \includegraphics[width=\linewidth]{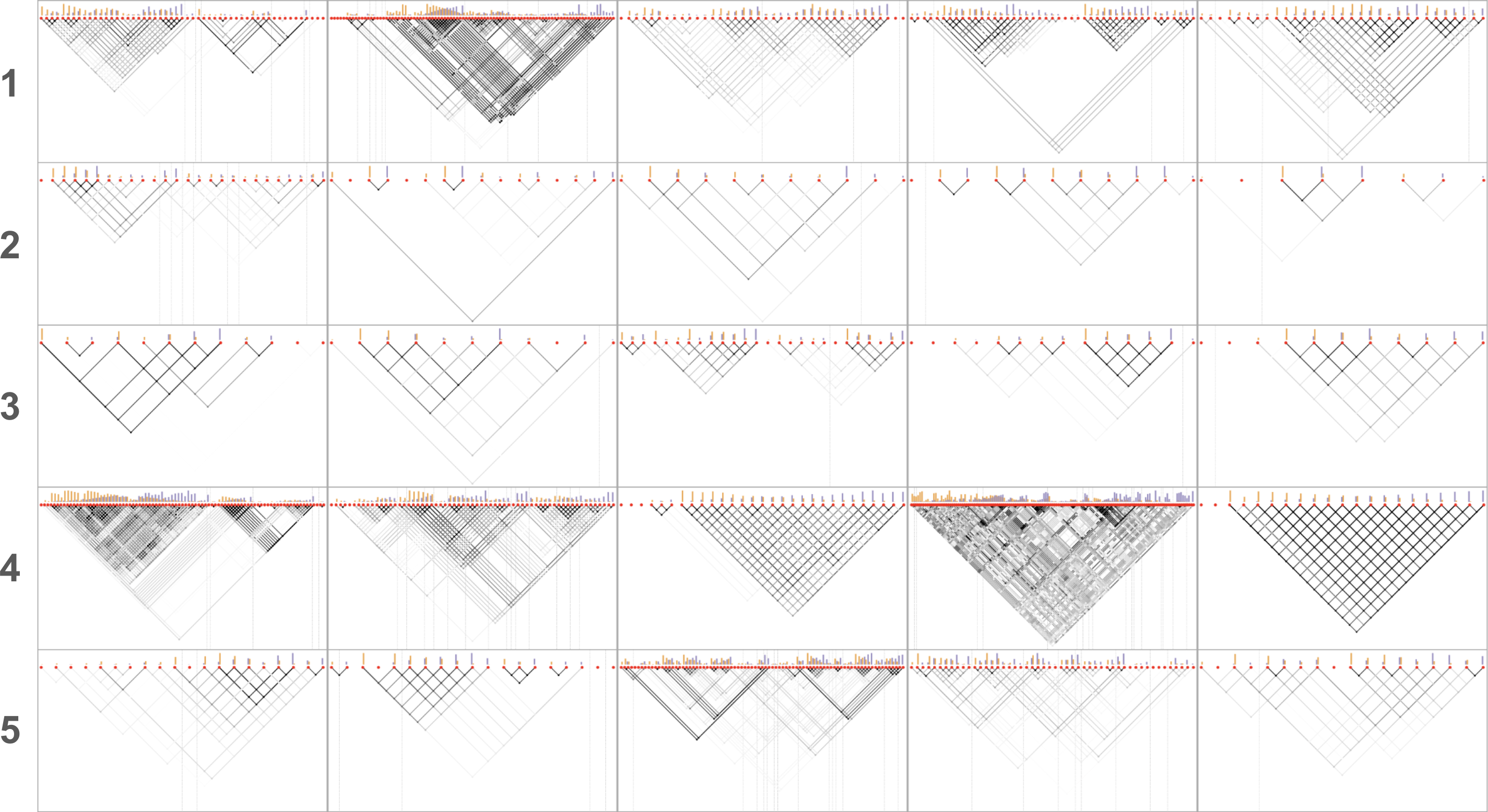}
    \caption{Five clusters of image prompting linkographs, five random examples per cluster.}
    \Description{Five rows of linkographs, labeled 1--5. Each row contains five linkographs. The linkographs in the first row have multiple clear chunks with a few longer-range links connecting these chunks. Those in the second row don't have many links; the links they do have tend to be either short-range or fairly weak. Those in the third row have at least one smallish web of strongly interconnected adjacent moves, but otherwise look a bit like those in the previous row. The linkographs in the fourth row are big and dense; some of them seem to be mostly taken up by a single giant web. Finally, those in the fifth row are zigzaggy and exhibit a mix of shorter-range and longer-range links.}
    \label{fig:Clusters}
\end{figure*}

Computing linkographic statistics on each trace also allows us to \emph{cluster} the traces, associating each user's activity with a distinguishable archetype. We first translate each trace into a \emph{signature vector} of three statistics: move count, link density index (LDI), and overall link entropy. We then normalize these values to z-scores and discard as an outlier any trace whose signature vector includes a z-score greater than 3; this excludes 59 traces, most of which are unusually long (consisting of hundreds of moves) and thus probably in need of partitioning for further analysis. We cluster the remaining traces' signature vectors using $k$-means clustering---as employed for design style clustering~\cite{DesignStyleClustering}---with $k = 5$.

The resulting clusters (Figure \ref{fig:Clusters}) can be described as follows:

\begin{enumerate}
\item Medium-long episodes featuring multiple distinct refinement webs \emph{(331 traces)}
\item Short-medium episodes involving mostly disconnected ideas \emph{(729 traces)}
\item Short-medium episodes containing a strong refinement web \emph{(273 traces)}
\item Long episodes of very densely interconnected moves, often featuring a very large refinement web \emph{(174 traces)}
\item Medium-long zigzaggy episodes with mostly short-range links but some longer-range links too \emph{(313 traces)}
\end{enumerate}

These clusters, although relatively rough, demonstrate the potential of \emph{designer modeling}~\cite{DesignerModeling} via automatically associating CST users with distinguishable user archetypes based on their linkographic activity patterns. For instance, members of cluster 1 often seem to repeatedly seek out an interesting region of image space, then gradually refine a single prompt until they converge on an image they like; members of cluster 2 mostly prompt for unrelated or largely unrelated concepts in batches of one or two prompts at a time; and members of cluster 4 often seem to have a single major \emph{fascination} and mostly prompt for images of this fascination.

Tracking the relative frequency of traces in each category might help guide design iteration on a deployed CST: for instance, a high frequency of short and low-connectivity episodes may indicate that users need additional scaffolding for goal formation. Meanwhile, cluster-based classification of individual users might allow in-the-moment adaptation of a CST's user interface to user-archetypal needs: for instance, users in cluster 5 may find more value than others in features involving the revisitation and recombination of past prompting subjects or themes.

\begin{figure*}
    \centering
    \begin{minipage}{0.32\textwidth}
        \centering
        \includegraphics[width=\linewidth]{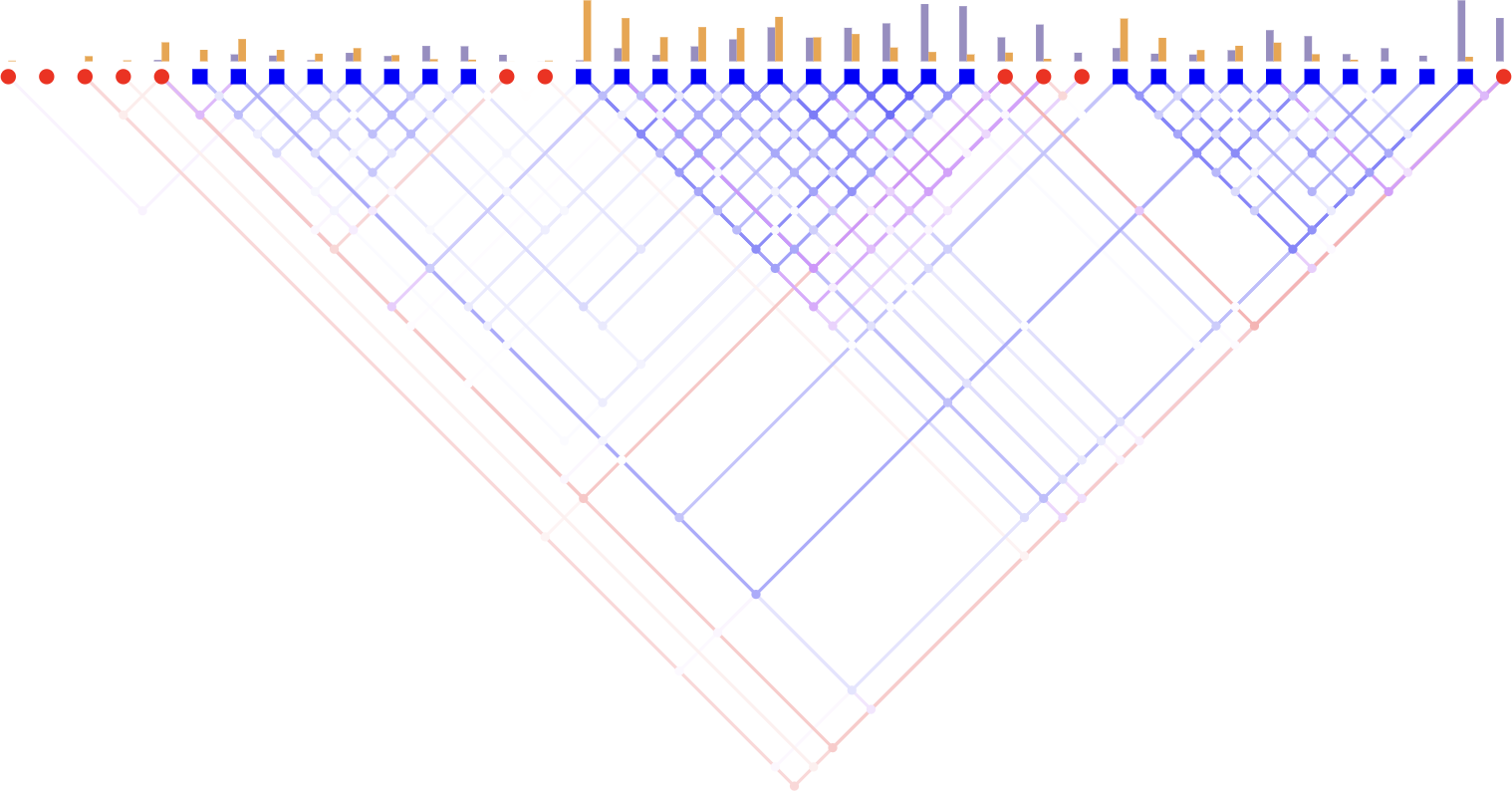}
        \Description{A linkograph that begins with a sequence of five human moves, mostly disconnected from one another. Next there's a loose chunk of machine moves, related incompletely to one another by relatively weak links. The user introduces a few more ideas that don't seem to take these machine moves into much account; then they prompt the machine again and get back a slightly larger cluster of more tightly interconnected ideas. They adapt a few of these, introduce a fresh idea of their own, and prompt once more, again yielding a tight cluster of machine moves in which the last few ideas seem a bit more loosely related than the rest. Finally, the user submits one more idea, seemingly uniting machine moves from multiple batches with one of the ideas they introduced at the very beginning of the episode.}
    \end{minipage}
    \hfill
    \begin{minipage}{0.32\textwidth}
        \centering
        \includegraphics[width=\linewidth]{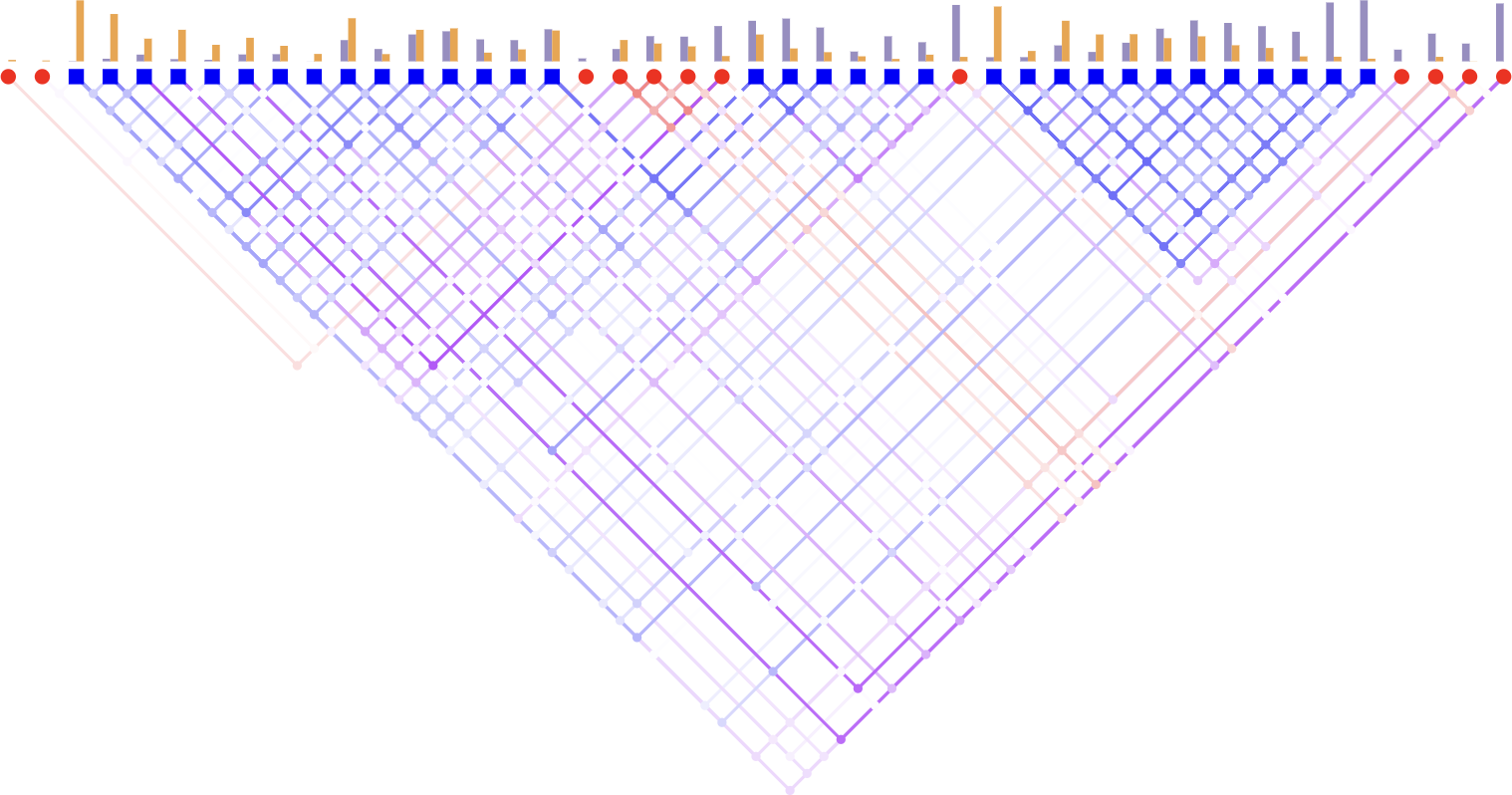}
        \Description{A linkograph similar to the previous, but with a different interleaving of human and machine moves. At the start there are three relatively disconnected human moves; these are followed by a dozen or so machine moves, which the user then begins converging into a shorter sequence of five human moves. The last human move in this five-move sequence seems very strongly linked to one of the machine moves in particular. Then there's another batch of (fewer, but more tightly interconnected) machine moves; a single human move building on these; another, larger and denser web of machine moves; and a few more human moves at the end, mostly linking back to machine moves from earlier in the trace.}
    \end{minipage}
    \hfill
    \begin{minipage}{0.32\textwidth}
        \centering
        \includegraphics[width=\linewidth]{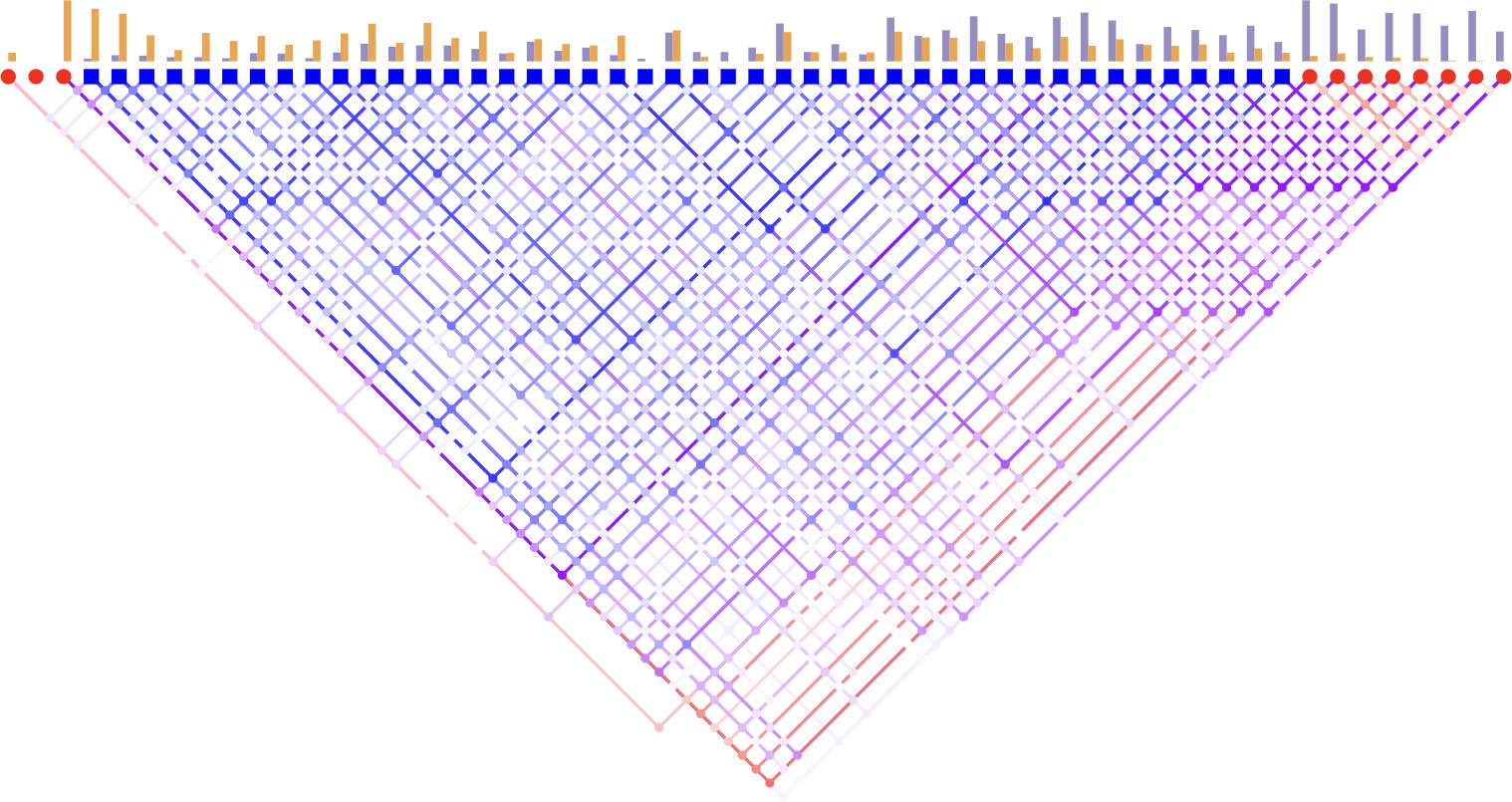}
        \Description{A third linkograph, similar to the other two but larger and more densely connected overall. The user starts off with three largely disconnected moves; then they invoke the machine several times, producing a progressively more densely interconnected long sequence of machine moves, probably on the order of three or four dozen in total. At the end, the user integrates several of these machine moves into a sequence of ten or so ideas.}
    \end{minipage}
    \caption{Three LLM-supported ideation linkographs demonstrating many of the archetypal patterns we observed in ideation traces. Early user ideas (\textcolor{red}{red}) are fairly diverse; the LLM (\textcolor{blue}{blue}) generates large batches of ideas at a time; LLM ideas tend to become more similar to one another as interaction proceeds; and inter-actor links (\textcolor{violet}{purple}) are dominated by user backlinks to LLM ideas, with users falling into a more curatorial or converging role as LLM output is introduced.}
    \label{fig:GiveAndTake}
\end{figure*}

\section{Analyzing LLM-Supported Ideation Sessions}

As a second case study, we use fuzzy linkography to analyze interaction dynamics between a user and a large language model in the context of open-ended divergent ideation, specifically focusing on what patterns of links can reveal about the creative ``give and take'' between two different actors (the human and machine). We examine 72 sequences of ideas introduced by a study participant and ChatGPT 3.5 in response to Product Improvement (PI) and Improbable Consequences (IC) ideation prompts. Data were collected in a prior study~\cite{HomogenizationEffects}, in which each participant generated sets of responses to two different prompts (one PI, one IC); participants were free to use ChatGPT however they preferred and spent eight minutes ideating on each prompt. Each episode of ideation thus consists of a sequence of ideas introduced during a single participant's engagement with a single ideation prompt.

In our LLM-supported ideation linkographs, ideas submitted by the participant via a text entry form are annotated as human design moves, and ideas generated as part of a participant-requested ChatGPT output are annotated as machine design moves. ChatGPT often generates blocks of many ideas at once, usually as a list of bullet points; we break these composite responses into multiple distinct machine moves, one per idea. Links are colorized to clarify influence dynamics: links between human moves are colored red, links between machine moves are colored blue, and links between human and machine moves are colored purple. 

\subsection{Inter-Actor Link Patterns}
\begin{table}
    \centering
    \begin{tabular}{r|ccc}
    & \textbf{Human} & \textbf{Machine (NC)} & \textbf{Machine (YC)} \\
    \hline
    \textbf{Human to...} & 0.0810 & 0.0701 & 0.1099 \\
    \textbf{Machine to...} & 0.0413 & 0.1055
    \end{tabular}
    \caption{Average density of backlinks from moves by the left actor to moves by the top actor, across LLM-supported ideation linkographs. ``Machine (NC)'' values exclude verbatim human copy-pastes of machine text from consideration; ``Machine (YC)'' values include them. 
    }
    \label{tab:BacklinkDensity}
\end{table}

\begin{figure}
    \centering
    \includegraphics[width=\linewidth]{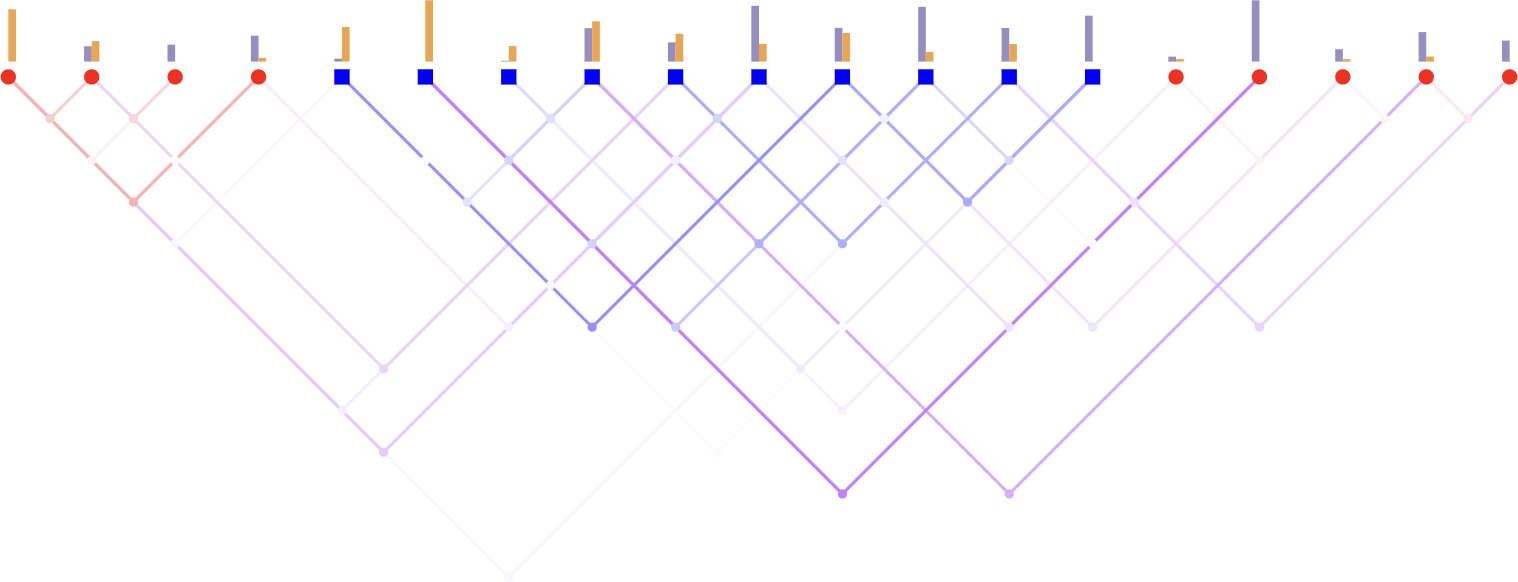}
    \caption{Even in LLM-supported ideation linkographs where user-introduced ideas start out relatively densely interconnected, LLM ideas reinforce these early user ideas, and the user clearly builds on LLM ideas in their own later ideation, we see little support for user fixation as a driver of the homogenization effect.}
    \Description{A linkograph of mixed human and machine moves. First there are four human moves, some of which are linked to one another by medium-strength or weak red links. Then there are ten machine moves, some of which are linked to one another a bit more densely by medium-strength blue links; a few of these are also linked by weak purple links to early human moves. Finally there are five more human moves, which are mostly disconnected from one another but each of which is linked to a different machine move by a weak or medium-strength purple link.}
    \label{fig:LimitedInfluence}
\end{figure}

The most immediately obvious pattern across LLM-supported ideation linkographs is that links within a single actor's ideas tend to dominate links between different actors' ideas. This is both visible in linkograph connectivity patterns (Figure \ref{fig:GiveAndTake}) and quantifiable in terms of backlink density (Table \ref{tab:BacklinkDensity}): with the exception of verbatim human copy-pastes of machine-generated text, human moves backlink more frequently and strongly to earlier human moves, and machine moves backlink more frequently and strongly to earlier machine moves.

LLM backlinks to human ideas seem especially rare. 51 of 72 total traces (70.83\%) feature a greater total backlink density from machine moves to human moves than vice versa; additionally, two traces (2.78\%, both very short outlier traces) feature no links at all between human and machine moves. Across all traces, the average density of backlinks from human to machine moves is nearly 2$\times$ that of backlinks from machine to human moves---nearly 3$\times$ if human copy-pastes of machine-generated text are included as human moves.

Broadly, these results suggests that ChatGPT 3.5 and its human users generally did not build as much on one another's ideas as they did on their own. Insofar as actors did influence one another, the machine influenced the human substantially more than the human influenced the machine.

Though this overall pattern of influence may seem counterintuitive at first (since the human is ultimately in control of the interaction and can choose how to prompt the machine), we believe it may reflect the fact that users rarely provided the LLM with complete information about their own ideas and the goals they were pursuing. Neither participants' frequent verbal remarks about machine output (some evaluative, some developmental) nor their submitted ideas were usually looped back into their conversations with ChatGPT---perhaps because of the extra effort needed to add ideas and commentary to the ChatGPT conversation, or perhaps because participants did not believe the machine could make effective use of this additional information. Regardless of the cause, we saw clear evidence for user \emph{underexpression} of creative intent~\cite{IntentElicitation}.

Interestingly, we did not see significant evidence in linkographs of individual \emph{machine critical moves} that strongly constrained human ideation (indicated by a particularly high forelink density toward human moves). Users often copy-pasted curated fragments of machine output and integrated or extended machine-introduced ideas, but we did not observe them becoming hung up on exploring the implications of any singular machine idea for the rest of the session (Figure \ref{fig:LimitedInfluence}). This lends further support to \citet{HomogenizationEffects}'s assessment that the homogenization effect in LLM-supported ideation is not a fixation effect.

\subsection{Other Interaction Dynamics}
\begin{figure}
    \centering
    \includegraphics[width=\linewidth]{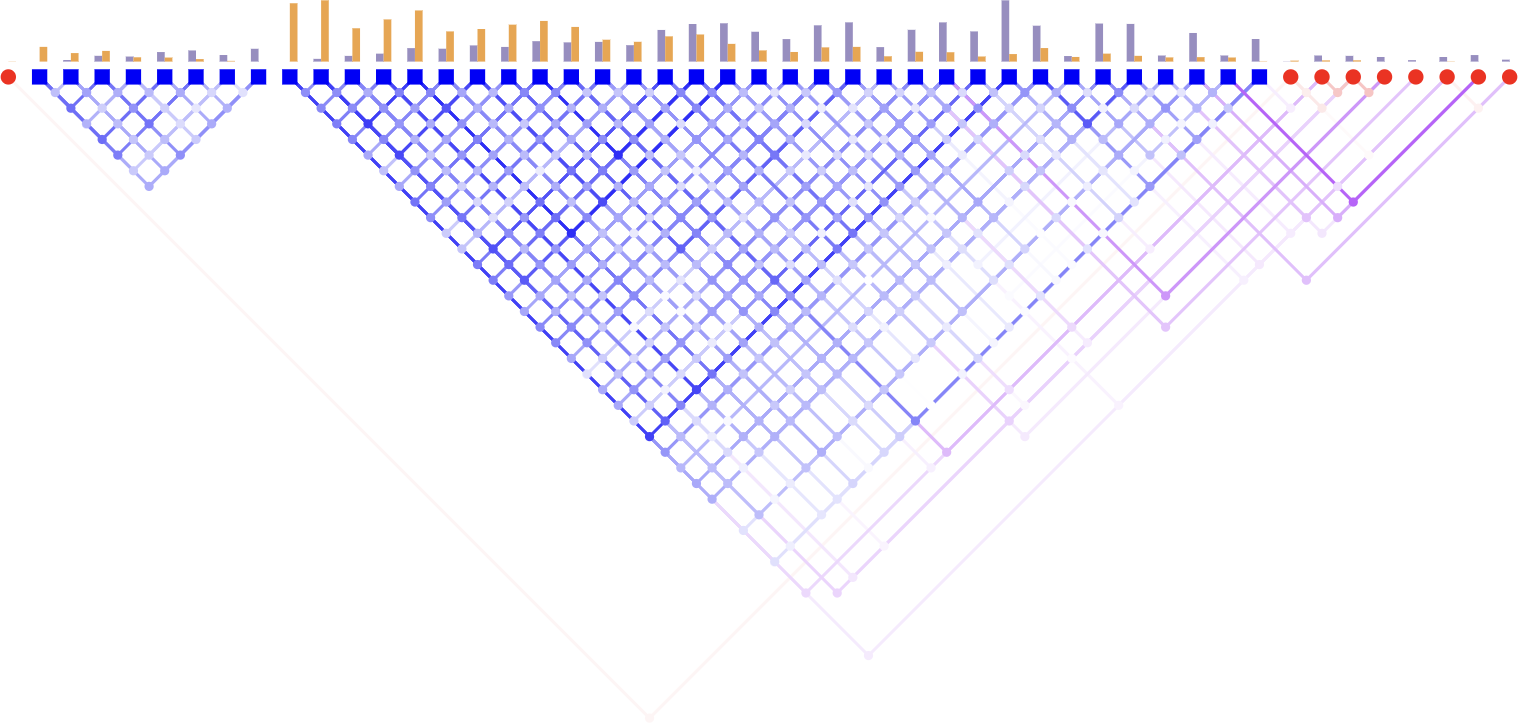}
    \caption{An LLM-supported ideation linkograph demonstrating multiple user invocations of the LLM in direct succession, with a different prompting strategy on each invocation. The first set of LLM ideas in particular are markedly disjoint from the later sets, and the user mostly discards these outputs when integrating LLM ideas into their own at the end.}
    \Description{A linkograph of mixed human and machine moves, with machine moves clearly predominating. A single human move is present at the beginning and doesn't seem to be linked to anything else. Then there's a densely interlinked web of six or seven machine moves, followed by a much longer and more densely interconnected sequence of machine moves that don't link back to the first web at all. The big chunk of machine moves opens with a very large and dense web but then seems to give way to a more moderately connected and smaller second web toward the end. Finally there are eight human moves at the very end of the graph, each building on one or more machine moves from the large chunk of machine moves in the middle of the graph.}
    \label{fig:MultiLLMCalls}
\end{figure}

We also observe several recurring patterns involving the evolution of linkographic structure over time. Most notably, LLM-introduced ideas tend to become more strongly linked to one another as conversation progresses. This trend becomes especially visible when the user requests multiple distinct ``batches'' of LLM ideas, as later idea batches more frequently form tightly interconnected webs rather than loosely interconnected chunks.

Progressive strengthening of links between LLM ideas may be attributed partly to users making more narrowly targeted queries to the LLM as the interaction progresses. For instance, a user may begin by simply asking the LLM for responses to the ideation prompt directly, then on follow-up requests ask it to take on a specific persona or focus on a particular aspect of ideas introduced before. In line with this explanation, we also observe that links \emph{between} LLM response batches tend to be weaker and less frequent than links within a single LLM response batch. Sometimes this even results in distinct batches of LLM-generated ideas being clearly separated in the linkograph despite the user not submitting any ideas of their own in between these LLM batches (e.g., Figure \ref{fig:MultiLLMCalls}).
Progressively greater interconnectedness between LLM ideas may also be partly due to the nature of LLMs as pattern-reproducing machines: as the context window becomes increasingly populated with specific keywords and ideation patterns, the LLM may become more likely to repeat itself or refer back to its own previous output.

User ideas also tend to be relatively disconnected from one another at first and to become gradually more connected as ideation progresses. This trend is in some ways more noticeable than for LLM ideas (because many ideation traces begin with several user ideas that are almost entirely disconnected from one another), and in some ways less noticeable (because user ideas never tend to become quite as strongly connected to one another as LLM ideas). In particular, prior to a user's first invocation of the LLM, their ideas tend to exhibit few links of any kind: study participants were asked to think divergently, and they initially seem to succeed at this task.

Nevertheless, once users first invoke the LLM, they tend to enter a mode of predominantly curating, merging, refining, or otherwise converging LLM suggestions. This leads to growing interconnectedness between user ideas. We attribute this shift partly to the LLM's tendency to generate large numbers of ideas in a single interaction turn: the ``one prompt in, many ideas out'' nature of human/LLM turntaking is very unlike the more balanced back-and-forth of typical human/human ideation, and may overwhelm the user with machine output to make sense of. To some extent, a user focus on making sense of machine output once it is available may reflect a normative view of the machine as a creative authority that gives the ``right answers'', in line with \citet{HomogenizationEffects}'s findings from qualitative analysis of user interviews.

\section{Analyzing Researcher Publication Histories}
As a third case study, and to illustrate the breadth of potential applications of our approach, we apply fuzzy linkography to the analysis of researcher publication histories as cataloged by Google Scholar. Research is not an activity carried out within a single CST, and a researcher's publication history may span many years, rather than the minutes or hours we typically associate with creative interaction dynamics; nevertheless, because publication activities are automatically transcribed as sequences of recognizable design moves, we can still apply fuzzy linkography to their analysis.

We examine the publication histories of 10 researchers, all either authors of this paper or well-known to us. This is a relatively small sample, but includes both academic and industry researchers at several distinct stages of career seniority (from PhD students up to associate professors) from a loosely related cluster of research fields centered on human-computer interaction. Publication histories, consisting of sequences of paper titles in publication order, were scraped from these researchers' public Google Scholar profiles; papers for which Google Scholar did not record any year of publication were excluded from our analysis, and we manually filtered out several Scholar-listed publications that did not constitute typical research contributions (e.g., complete proceedings volumes of workshops organized by the researcher in question; brief introductory notes at the front of conference proceedings volumes).

\begin{figure*}
    \centering
    \begin{minipage}{0.32\textwidth}
        \centering
        \includegraphics[width=\linewidth]{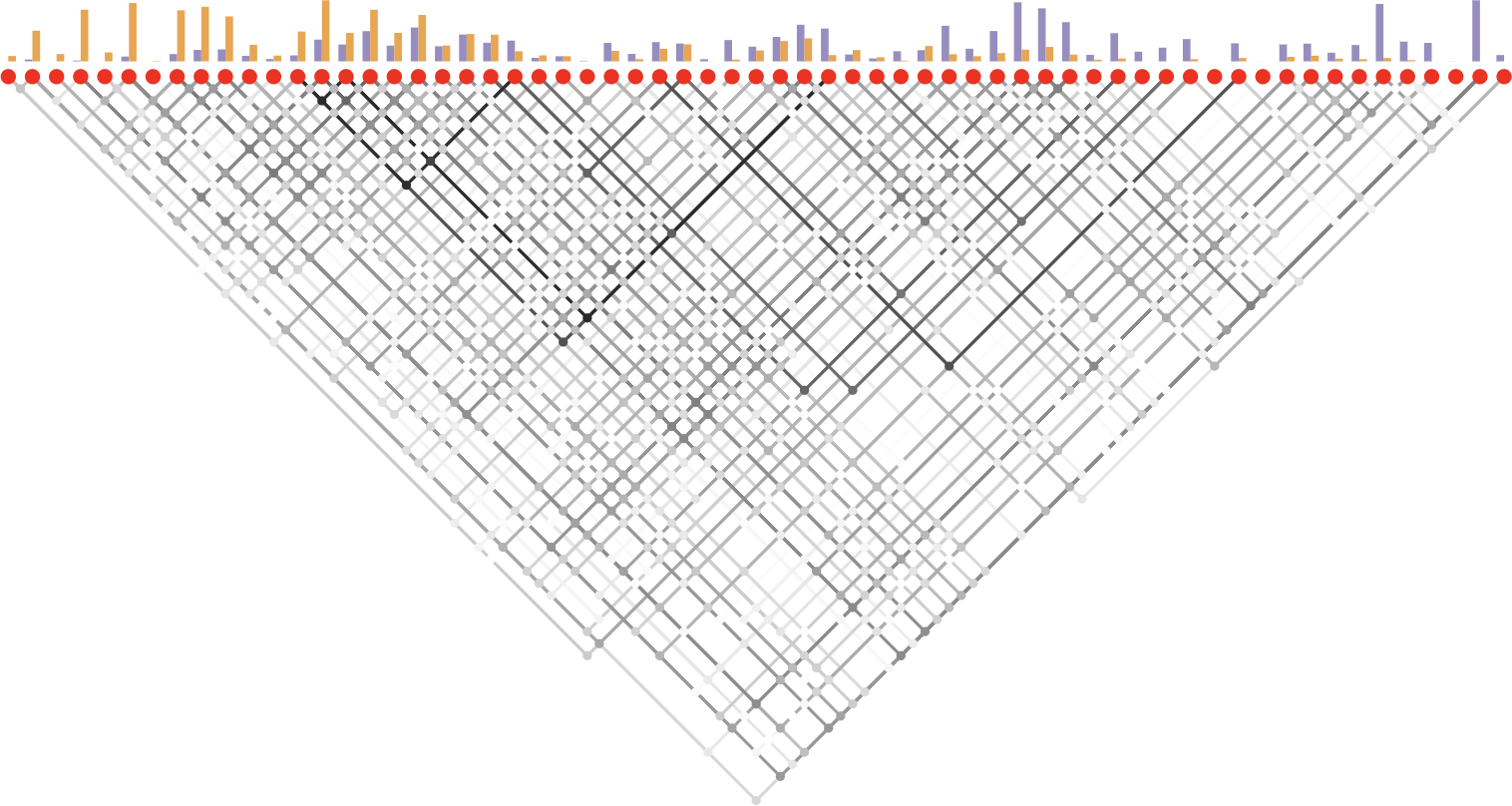}
        \Description{A linkograph consisting of many interrelated moves. Links of every length are mixed together, and most of the links are medium-strength. The overall impression is of a single giant chunk taking up the whole graph.}
    \end{minipage}
    \hfill
    \begin{minipage}{0.32\textwidth}
        \centering
        \includegraphics[width=\linewidth]{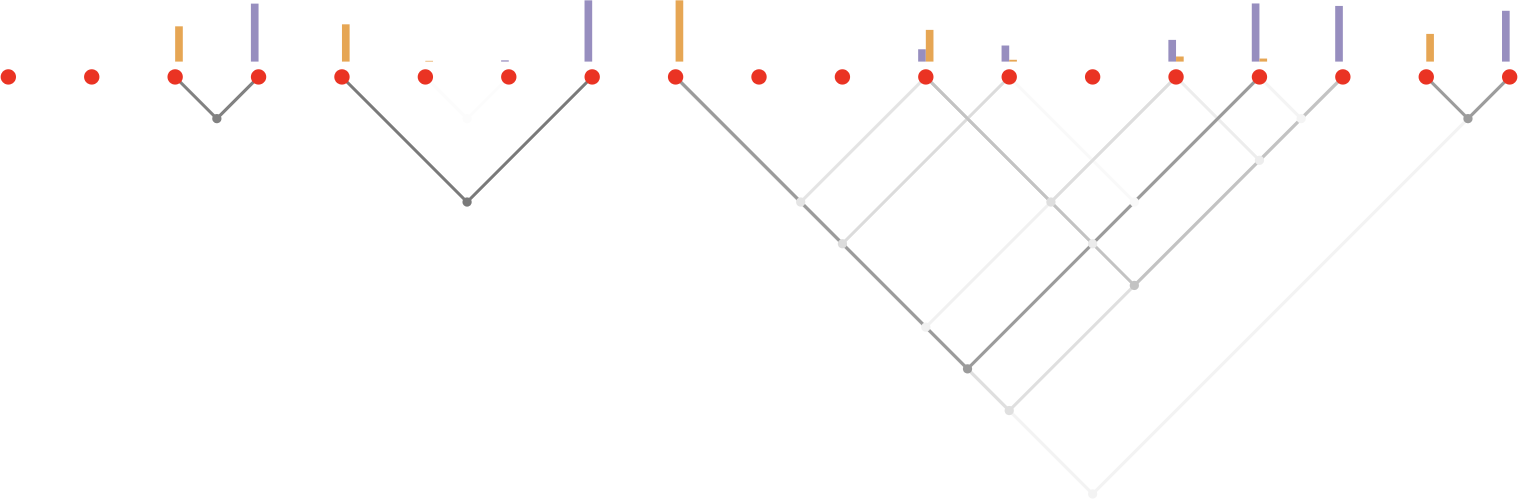}
        \Description{A linkograph of very sparsely interrelated moves. First there are two orphan moves; then there's a pair of moves linked to one another but nothing else. A four-move sequence follows, in which the first and last moves are linked only to one another, while the middle two moves are orphans. Then there's a largish chunk of maybe ten or so moves, with weak or medium-strength links interconnecting several but not all of the moves in this sequence. Finally there's one more pair of moves that are almost entirely disconnected from the previous chunk but connected to one another; the latter move in this pair is also connected very weakly to the move at the very beginning of the preceding chunk.}
    \end{minipage}
    \hfill
    \begin{minipage}{0.32\textwidth}
        \centering
        \includegraphics[width=\linewidth]{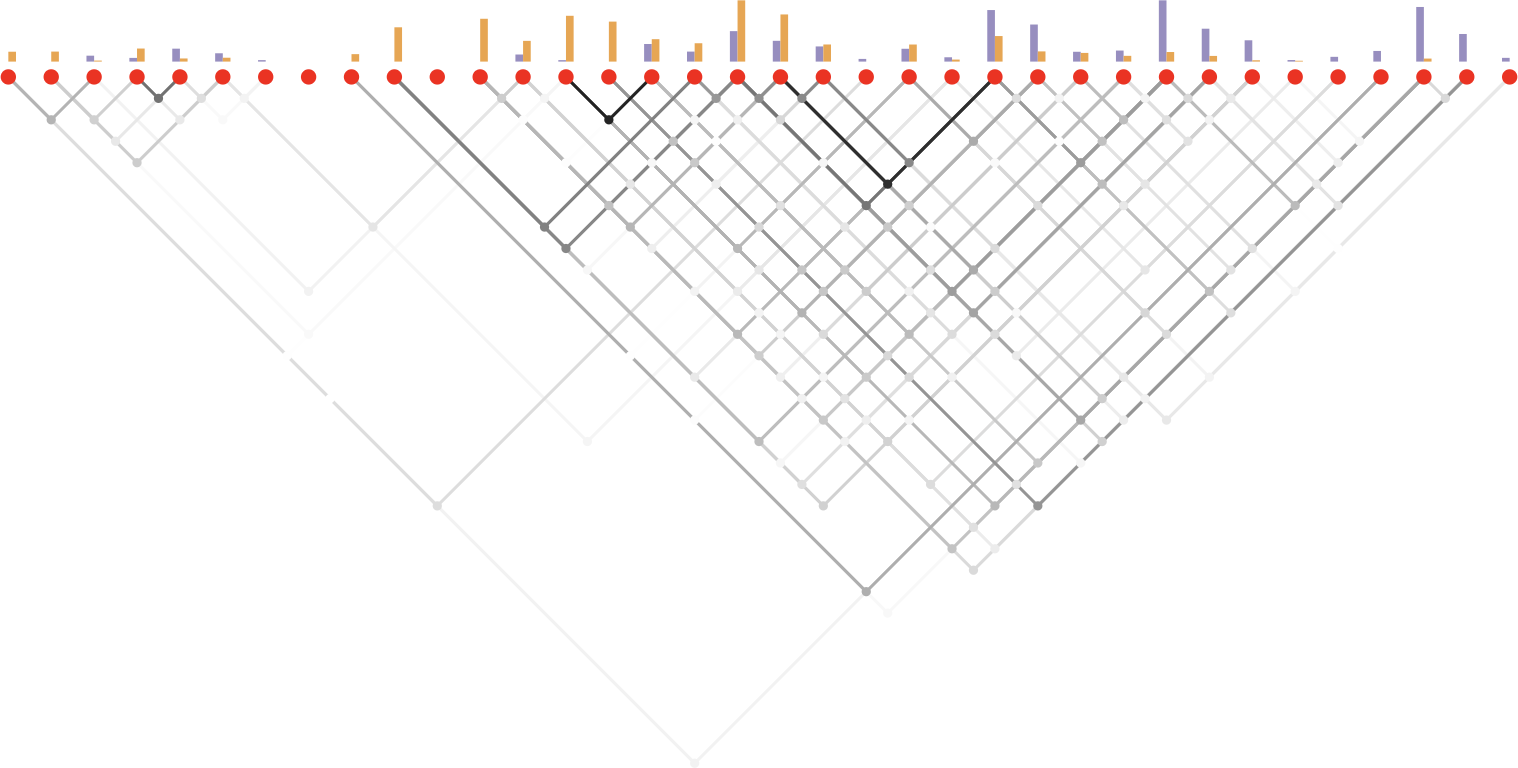}
        \Description{A linkograph featuring a chunk of six or so interrelated moves at the very beginning; a sequence of two or three orphan moves; and then a single large chunk of several dozen moves taking up most of the rest of the graph, with a handful of orphan moves interspersed. The two chunks in the graph are connected very weakly to one another by a handful of moves.}
    \end{minipage}
    \caption{Three publication history linkographs demonstrating three markedly different career shapes. The leftmost linkograph is most typical of the researchers in our sample (many of whom are still working on topics that are relatively close to their original research focus), but examples of researchers who have worked on a succession of largely unrelated topics and researchers who substantially ``pivoted'' before discovering their current primary research interest are also identifiable.}
    \label{fig:CareerShapes}
\end{figure*}

We begin by investigating whether significant papers can be automatically identified as \emph{critical moves} (CMs). The exact definition of a critical move varies from one study to the next, but moves with an especially high forelink or backlink count are often identified as critical; forelink CMs are interpreted as exerting a high degree of influence on future moves, and backlink CMs are interpreted as integrating implications of many previous moves~\cite{Linkography}. Therefore, in each publication history linkograph, we arbitrarily select the top three moves by forelink weight and the top three by backlink weight and mark these moves as critical.

Automatic CM identification seems to select papers that a human annotator might reasonably identify as critical roughly half the time. In approximately half of the graphs where a researcher's PhD thesis is present as a publication, the thesis is identified as a CM; in graphs where the researcher became known for a particular technical system early in their career, the first paper about this system is often identified as a CM (even when this paper is not the most cited paper about the system in question); and especially high-profile publications (e.g., ACM magazine articles about a researcher's work) also appear as CMs in several graphs.  Theses and high-profile articles may appear as CMs especially often because their titles tend to be \emph{general} rather than \emph{specific}, positioning them ``between'' many other articles by the same researcher in embedding space. However, roughly half of the papers identified as CMs by this procedure do not have any obvious significance to us.

Next, we identify several high-level linkographic structures that appeared in our dataset as distinguishable \emph{career shapes} (Figure \ref{fig:CareerShapes}). Most of the graphs we examined exhibit a broadly triangular shape in which links are almost equally prevalent at every possible pairwise move distance, suggesting a high degree of topic interrelatedness across a researcher's whole career. However, we also observe several outlier career shapes, including a ``wanderer'' structure (characterized by several highly disjoint clusters of papers published in different topic areas as the researcher repeatedly moved from one focus to another) and an ``early-career pivot'' structure (characterized by an early cluster of papers on one topic, followed by several one-off papers on largely unrelated topics and then a broad triangle shape indicating the development of a typical research focus).

Finally, we characterize the general structural patterns we observe in publication history linkographs. Most graphs we examined included at least one web structure: a sequence of several papers in which each paper is related to all the others. As in other contexts~\cite{UsingLinkography}, these webs seem to correspond to periods of ``mining out'' the consequences of a significant discovery made partway through the episode. However, no career consists entirely of webs: instead, webs mostly seem to be relatively small (on the order of 3-5 papers long) and separated by periods of more divergent exploration, presumably because veins of ``obvious'' follow-up work tend to become exhausted after a few papers in a row on similar topics by the same people at around the same time. Around half the linkographs we examined included a web structure near the end of the researcher's PhD period; where present, this generally seemed to correspond to the phase of ``cashing out'' dissertation work as publications after spending the earlier part of the PhD exploring different topics and ``building momentum'' to some extent.

On the contrary, we observe very few sawtooth structures---sequences of papers in which each one \emph{only} builds on its immediate predecessor---in the publication histories we examined. We hypothesize that this may be because individual papers are costly to produce, such that researchers are unlikely to spend much time on chains of follow-ups to papers that show no immediate sign of fitting neatly into their overall ``research narrative'': divergent papers tend to either become orphan moves, or end up forelinked to multiple downstream papers rather than just a single immediate successor. Potential sawtooths may also tend to be broken up by the alternating publication of papers from different projects or on different topics.

Overall, we found that our linkographic analysis of publication histories did not yield as many interesting insights as the other analyses we performed. This is as we expected: paper titles do not necessarily contain much information about the contents of each paper, so titles alone may not be readily interpretable as meaningful design moves; academic jargon may include terms that are closely related from the perspective of those ``in the know'', but hard for a generic text embedding model to pick up as similar; and academic careers are not especially similar to the shorter-term and more focused design situations that linkography is usually used to analyze, especially due to the individually costly nature of papers and the lack of information about research efforts that were never published (potentially ``failed'' design moves). Nevertheless, we include this analysis for three reasons:

\begin{enumerate}
    \item The high-level patterns we discovered strike us as interesting. In particular, the visibility within linkographs of distinctive ``career shapes'' and periods of ``mining out'' a compelling research idea both surprised us to some extent.
    \item This analysis was very inexpensive to perform, illustrating how rapid automatic construction of linkographs may shift linkography toward an approach that can be employed casually for low-cost exploration rather than one that requires a high degree of up-front investment to yield results.
    \item We suspect this analysis can be improved in the future, potentially by employing paper abstracts along with titles to form better design move summaries---or even embedding entire papers using a long-text embedding model.
\end{enumerate}

\section{Discussion}
\label{sec:discussion}

Potential applications of fuzzy linkography go well beyond those investigated here. For instance, large-scale application may be used to \emph{test linkographic hypotheses} about what linkographic patterns might indicate a successful episode of design or ideation (e.g., \cite{LinkographyTeamDesign}), or about how creativity functions as a process (e.g., \cite{LinkographicEvidence}). It is common in the linkography literature to associate quantitative measures on linkographs (e.g., entropy) with overall design process quality; associations of this sort may be easier to confirm or reject if linkographs are constructed at scale and measures on these linkographs directly compared against subjective ratings of creative experience~\cite{CSI,MICSI}, product quality, expressive communication~\cite{ExpressiveCommunication}, and so on. Meanwhile, theories of social creativity~\cite{SocialCreativityFischer,SocialCreativityWarr} and the differential effects of ``near'' versus ``far'' conceptual inspirations~\cite{DistantInspiration} (among others) may be especially amenable to testing via the multi-actor linkographic techniques we demonstrated in our study of LLM-supported ideation.



Realtime generation of linkographs opens the possibility of using linkography for \emph{reflective visualization}: displaying incrementally constructed linkographs to a design episode's participants while the episode is still unfolding, to give the participants an additional form of insight into their own creative process. Reflective visualization has proven helpful in co-creative contexts in the past~\cite{ConversationBalance}, and some CSTs are explicitly designed to promote reflection~\cite{ReflectiveCreators}, for instance to scaffold learning~\cite{AutomatedCritique,ImprovingCreativeFeedback,Shown}; the display of fuzzy linkographs to users may be especially impactful in this context.

Additionally, an AI-based CST that automatically and incrementally constructs a fuzzy linkograph during an episode of design could employ quantitative metrics on this linkograph to guide AI behavior. For instance, linkographic metrics could be used to gauge creative momentum~\cite{IntentElicitation}, to decide when would be a good time for the AI system to proactively intervene in a co-creative interaction~\cite{WhenToIntervene}; or to identify past moves that are as yet undeveloped, to focus proactive generation on unexplored potentialities.


Like homogenization analysis~\cite{LeveragingIdeasFromOthers,HomogenizationEffects}, fuzzy linkography can be applied to any artifacts that can be embedded in a semantic space---not just short texts but also images~\cite{CLIP,SigLIP}, longer texts~\cite{sturua2024jinaembeddingsv3multilingualembeddingstask, warner2024smarterbetterfasterlonger}, music~\cite{guo2023domainknowledge, huang2022mulan}, user interfaces~\cite{UIClip}, 3D models~\cite{3DModelEmbeddingA,3DModelEmbeddingB}, game states~\cite{GameMoments}, and likely others in the future. To support the use of alternative artifact types in future work, our implementation of fuzzy linkography includes support for the offline precalculation of links between moves. Future researchers can therefore employ their embedding model of choice for link inference and use our code for the final visualization of the calculated links.

For CSTs that produce more structured logs of user activity, it may also be possible to construct linkographs automatically without relying on an embedding model or other heuristic process to infer links between moves. When analyzing activity logs from a game design CST that automatically records what game entities are modified by each user action, for instance, it may be possible to establish clear binary links between moves if and only if they involve overlapping sets of entities. Similarly, linkographs of research publications might be constructed through analysis of citations or coauthorship links rather than textual similarity.

\section{Limitations and Future Work}
A fuzzy linkograph is fuzzy: it does not perfectly match the linkograph that a human annotator might produce for the same design episode. This is partly because even human annotators tend to disagree to some extent about how to construct linkographs from the same design moves~\cite{UsingLinkography}, but also partly because embedding models are imperfect proxies for human perceptions of design move relatedness. In particular, transformer-based sentence embedding models tend to place disproportionate emphasis on certain parts of speech~\cite{SentenceTransformerBias} and may overlook less obvious semantic relationships between design moves in some cases; for instance, a link was not inferred between ``not much money in the kitty'' and ``piggy bank'' in Figure \ref{fig:ExampleFuzzyLinkograph}, even though both involve currency. One direction for future work might involve formal comparison between human-constructed and machine-constructed linkographs to gauge the extent of their agreement.

\begin{figure*}
    \centering
    \includegraphics[width=\linewidth]{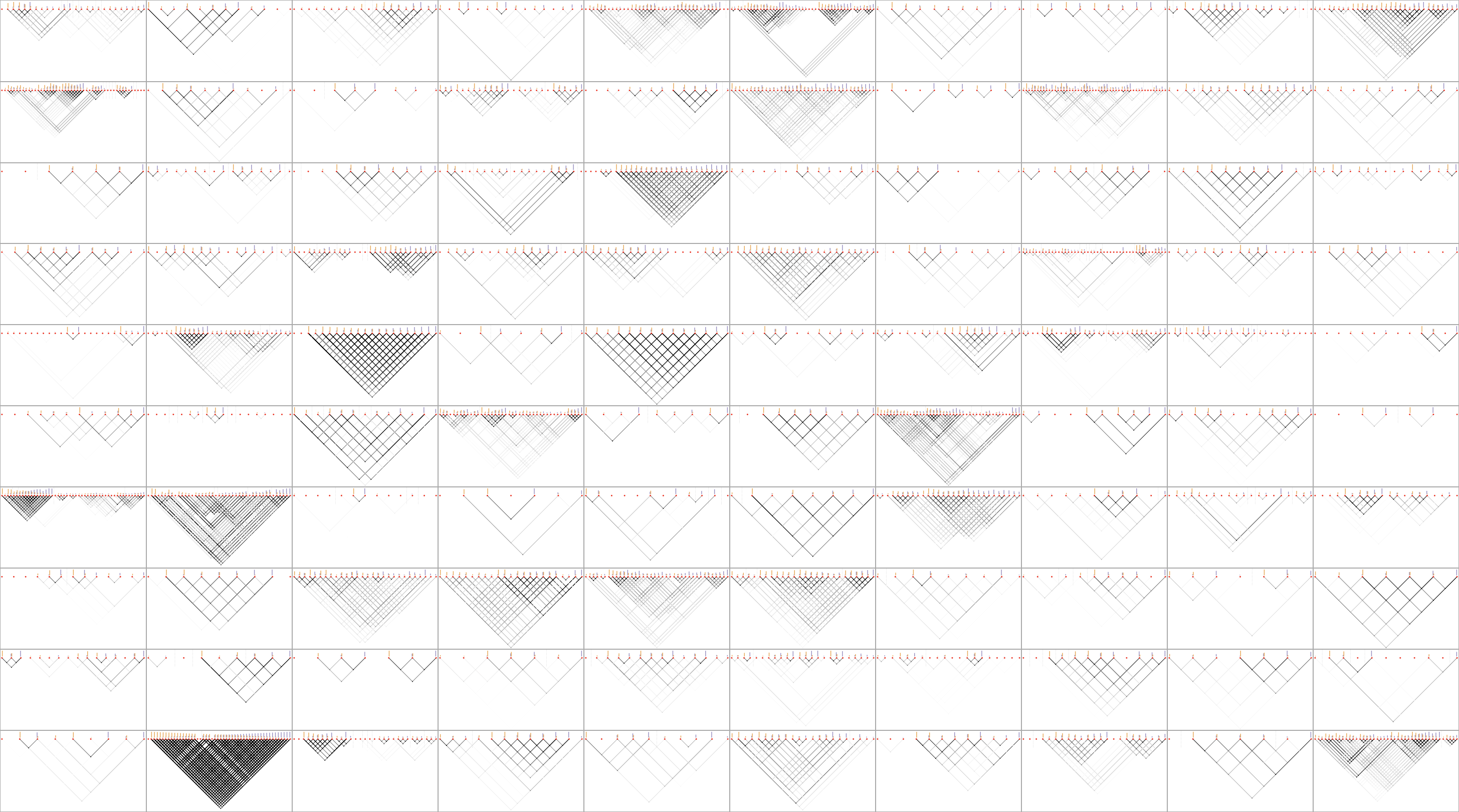}
    \caption{\emph{Linkographic abundance}:  fuzzy linkographs of 100 randomly selected image prompting traces, laid out as thumbnails. As new activity traces are logged, CST researchers might automatically generate linkographs for these traces and monitor the overall distribution, or use interesting-looking linkographs as jumping-off points for deeper investigation.}
    \Description{One hundred thumbnail-sized, structurally heterogenous linkographs packed into a single image. Some of them jump out more than others; to the bottom left there's one that looks like a single giant chunk except for a few entirely unrelated moves, and there are some other really dense ones too. When you spend longer looking at the thumbnails, more interesting patterns start to appear; there are some that have short-range and medium-range links but no long-range ones, some that have almost no links at all, some that contain clear zigzag shapes, and some that seem to get visibly denser or sparser over the course of the whole move sequence.}
    \label{fig:LinkographicAbundance}
\end{figure*}

To construct our linkographs, we used \texttt{all-MiniLM-L6-v2}~\cite{SentenceTransformers}---an open, general-purpose embedding model that has previously been validated against a human baseline for assessment of semantic similarity in creativity research~\cite{HomogenizationEffects}. However, better performance could likely be achieved with an embedding model specifically tuned on data from human-constructed linkographs. It may also be possible to use a large language model (or some alternative approach) rather than an embedding model to predict links between design moves; some automated creativity assessment pipelines seem to perform better when embeddings are replaced with an LLM that has been fine-tuned on task-specific data~\cite{BeyondSemanticDistance}, though this would likely increase cost per linkograph considerably.

Our reformulation of link entropy measures for fuzzy linkography is not necessarily the best one possible: in particular, by averaging continuous link strengths and treating the result as a probability of a binary link either existing or not existing, we arguably conflate association strength and probability of association in a way that could sometimes disguise meaningful differences between linkographs. To some extent this is an inherent risk of flattening complex graph structures down to summary values, and our entropy results on real data still seem to follow intuitive expectations---but it may nevertheless be desirable to invent an alternative quantitative proxy for design episode ``dynamism'' that yields comparatively higher values for fuzzy linkographs consisting of mixed weak and strong links. Similar criticisms have previously been leveled of the conventional approach to calculating link entropy in general; alternative formulations of entropy proposed in the past~\cite{CritiqueLinkEntropy,EntropyOfLinkography} may be adaptable to fuzzy linkography as well.

Linkography itself (especially when viewed as a visualization technique) does not necessarily scale well with trace length past a certain threshold: a linkograph containing more than a few dozen moves tends to become visually overwhelming. However, clustering or summarization of moves may permit condensation of longer traces to effectively shorter ones for linkographic purposes~\cite{LinkographyCondensation}, and quantitative linkographic metrics may still support trace clustering, critical move identification, and so on even at larger scales. Future work might involve evaluation of fuzzy linkograph interpretability---both with and without condensation techniques or other potential improvements applied---via formal user study.

\section{Conclusion}

We have introduced a technique for the automatic construction of \emph{fuzzy linkographs} from creative activity traces, including those collected naturally through the course of user interactions with digital creativity support tools. We have also demonstrated the application of this technique to a variety of creative domains, showcasing the versatility of our approach. Although the resulting linkographs are imperfect, they nevertheless function well as \emph{graphical summaries} of design episodes, surfacing high-level structural patterns in creative activity traces at a glance and serving as jumping-off points for deeper analysis. In closing, we would especially like to stress the potential for linkographic \emph{abundance} (Figure \ref{fig:LinkographicAbundance}) to enable new applications of linkography: whereas linkographs have up until this point been relatively scarce, the availability of low-cost approaches to linkography may greatly expand the potential audience for linkographic techniques, for instance by allowing CST developers to rapidly create linkographs for very large numbers of user activity traces as a window into broad usage trends.

\bibliographystyle{ACM-Reference-Format}
\bibliography{bibliography}


\end{document}